\documentclass[aps,twocolumn,prc,superscriptaddress,showpacs,nofootinbib,floatfix,amssymb,amsfonts,amsmath]{revtex4-1}
\usepackage{graphicx,colordvi}
\usepackage{amssymb,amsmath}
\usepackage{array}
\usepackage{calc}
\usepackage{ifthen}
\usepackage{epsfig}

\def\d{\hbox{d}}
\def\be{\begin{equation}}
\def\ee{\end{equation}}
\def\bea{\begin{eqnarray}}
\def\eea{\end{eqnarray}}
\def\l{\label}
\def\r{{\bf r}}
\def\p{{\bf p}}
\def\q{{\bf q}}

\def\om{\omega}
\def\Om{\Omega}

\def\hahat{\hat{H}}

\def\hahat0{\hat{H}_0}

\def\bs{\bigskip}
\def\ms{\medskip}

\def\cos{\hbox{cos}}
\def\sin{\hbox{sin}}

\def\exp{\hbox{exp}}

\def\Im{{\mbox {\rm Im}}}

\def\d{\hbox{d}}
\def\eps{\epsilon}
\def\epsd{\varepsilon}
\def\e{e}
\def\epsi{\mathcal{E}}
\def\siml{\hbox{\kern.1em \lower.6ex \hbox{$\sim$} \kern-1.12em
 \raise.6ex \hbox{$<$} \kern.1em}}
\def\simg{\hbox{\kern.1em \lower.6ex \hbox{$\sim$} \kern-1.12em
 \raise.6ex \hbox{$>$} \kern.1em}}

\begin{document}

\title{
SLOPE-DEPENDENT NUCLEAR-SYMMETRY ENERGY WITHIN THE EFFECTIVE SURFACE
APPROXIMATION}
\author{J.P. Blocki}
\affiliation{\it National Centre for Nuclear Research,
PL-00681 Warsaw, Poland}
\author{A.G. Magner \thanks{magner@kinr.kiev.ua}}
\affiliation{\it Institute for Nuclear Research,  03680 Kyiv, Ukraine}
\author{P. Ring}
\affiliation{\it Technical Munich University,  D-85747 Garching, Germany}

\vspace*{-1.05cm}

\begin{abstract}
The effective-surface approximation is extended taking into account
derivatives of the symmetry-energy density per particle with respect 
to the mean particle density. The isoscalar and isovector particle 
densities in this extended
 effective-surface approximation are derived. The improved expressions
of the surface
 symmetry energy, in particular, its surface tension coefficients
 in the sharp-edged proton-neutron asymmetric nuclei take into account
important gradient terms of the energy density functional.
For most Skyrme forces the surface symmetry-energy constants
and the corresponding neutron skins and  isovector stiffnesses
are calculated as functions of the Swiatecki derivative of
the nongradient term of the
symmetry-energy density per particle with respect to the isoscalar density.
Using the analytical isovector surface-energy constants
in the framework of the Fermi-liquid droplet model
we find energies and sum rules of the isovector 
giant-dipole resonance structure  in a reasonable agreement with
the experimental data, and they are compared with other 
theoretical approaches.

{\bf Keywords:} Nuclear binding energy, liquid droplet model,
extended Thomas-Fermi approach, nuclear surface energy,
symmetry energy, neutron skin thickness, isovector stiffness.
\date{\today }
\end{abstract}
\maketitle
PACS numbers: 21.10.Dr, 21.65.Cd, 21.60.Ev, 21.65.Ef

 \section{Introduction}

Explicit and accurate analytical expressions for the particle density
distributions within the nuclear effective-surface (ES)
approximation were obtained in Refs.\ \cite{strtyap,strmagbr,strmagden}.
They take advantage of the saturation properties of nuclear matter
in the narrow diffuse-edge region in finite heavy nuclei.
The ES is defined as the location of points with a maximal density gradient.
An orthogonal coordinate system related locally to the ES is specified by
the distance $\xi$ of a given point from the ES and the tangent coordinate
$\eta$ parallel to the ES. Using the nuclear energy-density functional
theory, one can simplify the variational condition derived from minimization of
the nuclear energy at some fixed integrals of motion in the $\xi,\eta$
coordinates
within the leptodermous approximation. In particular, in the extended
Thomas-Fermi
(ETF) approach \cite{brguehak}, this can be done in sufficiently heavy nuclei
for any fixed deformation using the expansion in a small parameter
$a/R \sim A^{-1/3} \ll 1$
where $a$ is of the order of the diffuse edge thickness of the nucleus,
$R$ is the mean curvature radius of the ES, and
$A$ is the number of nucleons. The accuracy of the ES approximation
in the ETF approach was checked \cite{strmagden}
without spin-orbit (SO) and asymmetry terms by comparing results
with those of Hartree-Fock (HF) and other ETF
models for some Skyrme forces. The ES approach \cite{strmagden} was also
extended by taking into account the SO and asymmetry 
effects \cite{magsangzh,BMV,BMRV}.

Solutions for the isoscalar and isovector
particle densities and energies in the ES (leptodermous)
approximation of the ETF approach were applied to analytical
calculations of the surface symmetry energy, the neutron skin and isovector
stiffness coefficient in the leading 
order of the parameter $a/R$~\cite{BMRV}. Our results are 
compared with older investigations
\cite{myswann69,myswnp80pr96,myswprc77,myswiat85} within the liquid droplet
model
(LDM) and with more recent works
\cite{vretenar1,vretenar2,ponomarev,danielewicz1,pearson,danielewicz2,danielewicz3,vinas1,vinas2,vinas4,kievPygmy,larsenprc2013,nester1,nester2,nester3,vinas5}.

The splitting of the isovector giant-dipole resonances into the main and 
satellite modes \cite{Endressplitting}
was obtained as a function of the
isovector surface-energy constant within the  Fermi-liquid droplet (FLD)
model \cite{kolmag,kolmagsh} in the ES approach.
The analytical expressions for the surface symmetry-energy constants 
have been tested by energies and sum rules
of the isovector
dipole resonances (IVDR) within the FLD model \cite{BMRPhysScr2014}
for some Skyrme forces neglecting derivatives of the nongradient
terms in the symmetry-energy density per particle with respect to
the mean particle
density.  The so called pygmy dipole resonances (PDR) below the main
IVDR peak as a different phenomenon
which might not be the result
of splitting of the IVDR were intensively discussed
in the literature 
\cite{vretenar1,vretenar2,ponomarev,kievPygmy,larsenprc2013,nester1,nester2,adrich,wieland,reviewpygmy1,reviewpygmy2}. 
They might have a different nature and actually are not related to
each other.
In the present work, we shall extend the variational effective-surface
method accounting for the derivatives introduced by Swiatecki and Myers
within the LDM \cite{myswann69} 
and apply it to the IVDR splitting. Some preliminary results were reported in
\cite{BMRPhysScr2015}.

In Sec.\ II, we give an outlook of the basic points of the ES approximation
within the density functional theory. The
main results for the isoscalar and isovector
particle densities are presented in Sec.\ III 
emphasizing derivatives of the symmetry energy density per particle.
Section IV is devoted to analytical derivations of the
symmetry energy in terms of the surface energy coefficient, the neutron skin
thickness
and the isovector stiffness including these derivatives.
Sections V and VI are devoted to the collective dynamical description
of the IVDR structure in terms of the response functions and transition
densities. Discussions
of the results are given in
Sec.\ VII and summarized in Sec.\ VIII.
Some details of our calculations are presented in Appendixes A and B.

\section{Symmetry energy and particle densities}

We start with the nuclear energy $E$ as a functional of
the isoscalar ($\rho_{+}$) and isovector  ($\rho_{-}$) densities
$\rho_{\pm}=\rho_n \pm \rho_p$ in the local density approach
\cite{brguehak,chaban,reinhard,reinhardSV,bender,revstonerein,reinhardSV,ehnazarrrein,pastore,jmeyer}:
\be\l{energy}
E=\int \d \r\; \rho_{+}\epsi\left(\rho_{+},\rho_{-}\right),
\ee
where
$\epsi\left(\rho_{+},\rho_{-}\right)$
is the energy density per particle,
\bea
&&\epsi\left(\rho_{+},\rho_{-}\right) =
- b^{}_V + J I^2
+\epsd_{+}(\rho_{+}) + \epsd_{-}(\rho_{+},\rho_{-})
+\quad \nonumber\\
&&+
\left(\mathcal{C}_{+}/\rho_{+} + {\cal D}_{+} \right)
\left(\nabla \rho_{+}\right)^2
+ \left(\mathcal{C}_{-}/\rho_{+} + {\cal D}_{-} \right)
\left(\nabla \rho_{-}\right)^2.
\label{enerdenin}
\eea
Here, $b^{}_V \approx$ 16 MeV is the separation energy of a particle,
$J \approx $ 30 MeV is the main volume symmetry-energy constant of
infinite nuclear matter, and $~I=(N-Z)/A$ is the asymmetry parameter;
$~N=\int \d \r \rho_n(\r)$ and $~Z=\int \d \r \rho_p(\r)$ are
the neutron and proton numbers, and $~A=N+Z$. These constants determine
the first two terms of the volume energy. The last four terms
are surface terms. The first two terms are independent of
the gradients of the particle densities and the last two
depend on these gradients. For the first surface
term independent of the gradients,
$\epsd_{+}$, one obtains
\be\l{epsilonplus0}
\epsd_{+}(\rho_{+})=\frac{K_{+}}{18}\;
e_{+}\left[\eps(w_{+})\right],
\ee
where $K_{+}\approx 215-245$ MeV (see Table I) is the isoscalar
in-compressibility modulus of symmetric nuclear matter,
$w_{+}$ is the dimensionless isoscalar particle density,
$w_{+}=\rho_{+}/\overline{\rho}$ and
\be\l{epsilonplus}
e_{+}\left[\eps\left(w_{+}\right)\right]=9 \eps^2 +
I^2\left[\mathcal{S}_{\rm sym}(\eps)-J\right]/K_{+} 
\ee
with
\be\l{eps}
\eps=\frac{\overline{\rho}-\rho_{+}}{3\overline{\rho}}=\frac{1-w_{+}}{3}.
\ee
$\eps$ is the small parameter in the expansion,
\be\l{symen}
\mathcal{S}_{\rm sym}(\eps)= J - L \eps +\frac{K_{-}}{2} \eps^2 + ...,
\ee
around the particle density of infinite nuclear matter
$\overline{\rho}=3/4\pi r_0^3 \approx$ 0.16 fm$^{-3}$ and $r^{}_0$
is the commonly accepted constant in the $A^{1/3}$ dependence of a mean radius.
Several other quantities, which were
introduced by Myers and Swiatecki \cite{myswann69}, will be explained below.
The derivative corrections of $\mathcal{S}_{\rm sym}(\eps)$ in Eq.\ 
 (\ref{symen})
were neglected in our previous calculations \cite{BMRV}.
The next isovector surface term $\epsd_{-}(\rho_{+},\rho_{-})$
can be defined
through the same function $\mathcal{S}_{\rm sym}(\eps)$ in Eq.\ (\ref{symen}):
\be\l{surfsymen}
\epsd_{-}\left(\rho_{+},\rho_{-}\right)=
\mathcal{S}_{\rm sym}(\eps)\;
\left(\frac{\rho_{-}}{\rho_{+}}\right)^2 - J I^2.
\ee
For the first and second derivatives of $\mathcal{S}_{\rm sym}(\eps)$ with
respect to $\eps$
one can take in Eq.\ (\ref{symen})  the values
$L \approx 20 \div 120$ MeV and, even less known,
$K_{-}$ \cite{PR2008,vinas1,vinas5}.
The constants $\mathcal{C}_{\pm}$ and $\mathcal{D}_{\pm}$ in
Eq.\ (\ref{enerdenin}) are defined by the parameters of the Skyrme forces
\cite{brguehak,bender,chaban,reinhardSV,pastore},
\bea\l{Cpm}
\mathcal{C}_{+}&=&
\frac{1}{12} \left(t_1 - \frac{25}{12} t_2 - \frac{5}{3} t_2 x_2\right),\\
\mathcal{C}_{-}&=&-\frac{t_1}{48}\left(1 + \frac{5}{2}x_1\right) -
\frac{t_2}{36} \left(1 + \frac{19}{8} x_2\right).\nonumber
\eea
The isoscalar SO gradient terms in (\ref{enerdenin}) are defined
with a constant:
$\mathcal{D}_{+} = -9m W_0^2/16 \hbar^2$, where
$W_0 \approx$ 100--130 MeV$\cdot$fm$^{5}$ and
$m$ is the nucleon mass.
The constant $\mathcal{D}_{-}$ is usually relatively
small and will be neglected below for simplicity.
Equation (\ref{enerdenin})
can be applied in a semiclassical approximation for a realistic
Skyrme force \cite{chaban,reinhardSV,bender,revstonerein,ehnazarrrein},
in particular by neglecting higher $\hbar$ corrections
in the ETF kinetic energy \cite{brguehak,strmagbr,strmagden}
and also Coulomb terms.
All of them easily were taken into account
\cite{strtyap,magsangzh} neglecting relatively small Coulomb exchange terms.
Such exchange terms can be calculated numerically in extended Slater
approximations~\cite{GU-HQ2013_PRC87-041301}.

The energy density per particle in Eq.\ (\ref{enerdenin}) contains
the first two volume terms and surface components
including the new $L$ and $K_{-}$ derivative corrections
$\epsd_{-}$ (in contrast to Ref.\ \cite{BMRV}) and also the isoscalar
and isovector  density gradients.
Both are important for finite nuclear systems.
These gradient terms together with the other surface components
in the energy density within the ES approximation
are responsible for the surface tension in finite nuclei.

As usual, we minimize the energy $E$ under the constraints
of fixed particle number $A=\int \d \r\; \rho_{+}(\r)$ and
 neutron excess $N-Z= \int \d \r\; \rho_{-}(\r)$ using
the Lagrange multipliers $\lambda_{+}$ and $\lambda_{-}$, the
isoscalar and isovector chemical-potential surface corrections
(see Appendix A).
Taking also into account additional deformation constraints
(like the quadrupole moment), our approach can be applied for any deformation
parameter of the nuclear surface, if its diffuseness $a$ is small
with respect to the curvature radius $R$.
Approximate analytical expressions of the binding energy will be obtained
at least up to order $A^{2/3}$. To satisfy the condition of particle
number conservation
with the required accuracy we account for relatively small
surface corrections ($\propto a/R \sim A^{-1/3}$ in first order)
to the leading terms in the Lagrange multipliers
 \cite{strmagbr,strmagden,magsangzh,BMRV}
(see Appendix B). We take into account explicitly the diffuseness
of the particle density distributions. Solutions of the
variational Lagrange equations can be derived analytically
for the isoscalar and isovector
surface tension coefficients (energy constants), instead of the
phenomenological constants of the standard LDM \cite{myswann69}
(the neutron and proton particle densities were considered
earlier to be distributions with a strictly sharp edge).

\section{Extended isoscalar and isovector densities}

For the isoscalar particle density, $w =\rho_{+}/\overline{\rho}$, one has
up to the leading terms in the leptodermous
parameter  $a/R$ the usual first-order
differential Lagrange equation~\cite{strmagden,magsangzh,BMRV}.
 Integrating this equation, one finds the solution:
\be\l{ysolplus}
x=-\int\limits_{w_r}^{w}\d y\; \sqrt{\frac{1 +\beta y}{y\;e_{+}[\eps(y)]}}\;,
\qquad
x=\frac{\xi}{a},
\ee
for $x < x(w=0)$ and $w=0$ for $x \geq x(w=0)$, where $x(w=0)$
is the turning point.
$\beta=\mathcal{D}_{+} \overline{\rho}/\mathcal{C}_{+}$ is the dimensionless SO
parameter, see Eq.\ (\ref{epsilonplus}) for $e_{+}[\eps(y)]$
(for convenience we often omit the lower index ``$+$'' in \\
$w_{+}$ ).
For $w_r=w(x=0)$, one has the boundary condition, $\d^2 w(x)/\d x^2=0$
at the ES ($x=0$):
\be\l{boundcond}
e_{+}[\eps(w_r)] + w_r(1 +\beta w_r)
\left[\frac{\d e_{+}[\eps(w)]}{\d w}\right]_{w=w_r}=0\;.
\ee

In Eq.\ (\ref{ysolplus}), $a \approx 0.5-0.6$ fm is the
diffuseness parameter \cite{BMRV},
\be\l{adif}
a=\sqrt{\frac{\mathcal{C}_{+} \overline{\rho} K_{+}}{30 b_V^2}},
\ee
found from the asymptotic behavior of the particle density,
$w \sim \exp(-\xi/a)$ for large $\xi$ ($\xi \gg a$).

As shown in Refs.\ \cite{strmagden,magsangzh},  the influence of the
 semiclassical $\hbar$ corrections (related to the ETF
kinetic energy) to $w(x)$ is negligibly small everywhere,
except for the quantum tail outside
the nucleus ($x \simg 1$). Therefore, all these corrections were neglected in
Eq.\ (\ref{enerdenin}).
With a good convergence of the expansion of the
$e_{+}[\eps(y)]$ in powers of $1-y$
 up to the leading quadratic term \cite{strmagden,magsangzh}
and small $I^2$ corrections in Eq.\ (\ref{epsilonplus}),
$e=(1-y)^2$,
one finds analytical solutions of Eq.\ (\ref{ysolplus}) in terms of the
algebraic, trigonometric and
logarithmic functions \cite{BMRV}.
For $\beta=0$ (i.e. without SO terms), it simplifies to
the solution $w(x)=\tanh^2\left[(x-x_0)/2\right]$ for
$x\leq x_0=2{\rm arctanh}(1/\sqrt{3})$
and zero for $x$ outside the nucleus ($x>x_0$).

After simple transformations of the isovector Lagrange equation
(\ref{lagrangeqminus}) one similarly finds up to the leading term in $a/R$  in the ES approximation
for the isovector density, $w_{-}(x)=\rho_{-}/(\overline{\rho}I)$, the
equation and the boundary condition (\ref{yeq0minus}).
The analytical solution $w_{-}=w \cos[\psi(w)]$
can be obtained through the expansion (\ref{powser}) of $\psi$ in powers of
\be\l{csymwt}
\gamma(w)=\frac{3\eps}{c_{\rm sym}},\qquad
c_{\rm sym}=a \;\sqrt{\frac{J}{\overline{\rho}\;
\vert\mathcal{C}_{-}\vert}}\;.
\ee
Expanding up to the second order in $\gamma$ one obtains (see Appendix A)
%
%\bea\l{ysolminus}
\be\l{ysolminus}
w_{-}=  w\;\cos\left[\psi(w)\right]
\approx
w \left(1- \frac{\psi^2(w)}{2} + ...
\right),
\ee
%\eea
%
with
\be\l{psi2}
\psi(w)=\frac{\gamma(w)}{\sqrt{1+\beta}}\;
\left[1 + \widetilde{c} \gamma(w) + ...\right],
\ee
\be\l{ctilde}
\widetilde{c}=\frac{\beta c_{\rm sym}^2 + 2 +
c_{\rm sym}^2 L (1+\beta)/(3J)}{2 c_{\rm sym}(1+\beta)},
\ee
[see also the constant $c_3$ (Appendix A) at higher (third) order corrections].
The constant $\widetilde{c}$
[Eq.\ (\ref{ctilde})] for the
isovector solutions $w_{-}$, Eq.\ (\ref{ysolminus}), is modified with respect to
Ref.\ \cite{BMRV} in two aspects. In addition to
the $L$ dependence there are also higher order terms from a
nonlinear equation (\ref{ueqminus}) for $\psi(w)$ (Appendix A).
Notice also that $w_{-}$ depends on $L$ in second order
in $\gamma$ but it is independent of $K_{-}$ at this order (Appendix A).

In Fig.\ \ref{fig1}, the $L$ dependence of the
function $w_{-}(x)$ is shown within the total interval from
$L=0$ to $L=100$  MeV
\cite{vinas2} and it is  compared to
that of the density $w(x)$ for the SLy5* force as a typical example.
As  shown in Fig.\ \ref{fig2} in a
larger (logarithmic) scale,
one observes notable differences in the isovector densities $w_{-}$ derived
from different Skyrme forces \cite{chaban,reinhardSV}
within the edge diffuseness. All these calculations
have been done with the finite
proper value of the slope parameter $L$. For
SLy forces this value is taken from Ref.\
\cite{jmeyer}, for $SGII$ from Ref.\ \cite{vinas2} and for others from
Ref.\ \cite{reinhardSV} (Table I).  
As shown below, this is in particular important
for calculations of the neutron skin of nuclei.
Notice that, with the precision of line thickness,
our results are almost the same
taking approximately $L=50$ MeV for SLy5* and $L=60$ MeV for SVsym32.
Note also that, up to second order in the small parameter $\gamma$,
the isovector particle density $w_{-}$ in Eq. (\ref{ysolminus})
does not depend on the symmetry-energy in-compressibility $K_{-}$.
The $K_{-}$ dependence appears only at higher (third) order terms in the
expansion in $\gamma$ (Appendix A).
Therefore, as a first step of the iteration procedure, it is possible to study
first the main slope effects of $L$ neglecting small $I^2$ corrections to the
isoscalar particle density $w_{+}$ (\ref{ysolplus}) through $e_{+}$
(\ref{epsilonplus}).
Then, we may study more precisely the effect of the second derivatives
$K_{-}$ taking into account higher order terms.

We emphasize that the dimensionless densities, $w(x)$
[see Eq.\ (\ref{ysolplus}) and Ref.\ \cite{BMRV} ] and $w_{-}(x)$
(\ref{ysolminus}), shown in Figs.\ \ref{fig1} and \ref{fig2}, were
obtained in leading ES approximation ($a/R \ll 1$) as functions of
specific combinations
of Skyrme force parameters like $\beta$ and $ c_{\rm sym}$
[Eq.\ (\ref{csymwt})] accounting for the $L$-dependence
 [Eq.\ (\ref{ctilde})].
These densities are at the leading order in the leptodermous parameter
$a/R$ approximately universal functions, independent of the
properties of the specific nucleus.
It yields largely the local density distributions in the
normal-to-ES direction $\xi$
with the correct asymptotic behavior outside of the deformed
ES layer  at $a/R \ll 1$,
as is the case for semi-infinite nuclear matter.
Therefore, at the dominating order,
 the particle densities $w_{\pm}$
are universal distributions independent
of the specific properties of nuclei while higher order corrections
to the densities $w_{\pm}$
depend on the specific macroscopic properties of nuclei.

\section{Isovector energy and stiffness}

The nuclear energy $E$ [Eq.\ (\ref{energy})]
in the improved ES approximation (Appendix B) is split into volume and
surface terms \cite{BMRV},
\be\l{EvEs}
E \approx -b^{}_V\; A + J (N-Z)^2/A + E_S\;.
\ee
For the surface energy $E_S$ one obtains
\be\l{Es}
E_S = E_S^{(+)} + E_S^{(-)}
\ee
with the
isoscalar (+) and isovector (-) surface components:
\be\l{Espm}
 E_S^{(\pm)}= b_S^{(\pm)} \frac{\mathcal{S}}{4\pi r_0^2},
\ee
where $\mathcal{S}$ is the surface area of the ES, $b_{S}^{(\pm)}$
are the
isoscalar $(+)$ and isovector $(-)$ surface-energy constants,
\be\l{sigma}
b_S^{(\pm)} \approx 8 \pi r_0^2 \mathcal{C}_{\pm}
\int_{-\infty}^{\infty} \d \xi\;
\left(1 + \frac{\mathcal{D}_{\pm}}{\mathcal{C}_{\pm}} \rho_{+}\right)
\left(\frac{\partial \rho_{\pm}}{\partial \xi}\right)^2.
\ee
These constants are proportional to the corresponding surface tension
coefficients
$\sigma_{\pm}=b_S^{(\pm)}/(4 \pi r_0^2)$ through the
solutions (\ref{ysolplus}) and (\ref{ysolminus})
for $\rho_{\pm}(\xi)$ which can be taken into account in leading
order of $a/R$ (Appendix B). These
coefficients $\sigma_{\pm}$ are the same as found in the
expressions for the capillary pressures of the macroscopic
boundary conditions [see Ref.\ \cite{BMRV} with new values
$\epsd_{\pm}$ modified
by $L$ and $K_{-}$ derivative corrections of Eqs.\ (\ref{epsilonplus}) and (\ref{surfsymen})].
Within the improved ES approximation where also
higher order corrections in the small parameter $a/R$
are taken into account, we derived in Ref.\ \cite{BMRV} equations for the
nuclear surface itself (see also Refs.\ \cite{strmagbr,strmagden,magsangzh}).
For more exact isoscalar and isovector particle densities we account for the
main terms in the next order of the parameter $a/R$
in the Lagrange equations (see Eq.\ (\ref{lagrangeqminus}) for the isovector and
Refs.\ \cite{strmagbr,strmagden,magsangzh} for the isoscalar case).
Multiplying these equations by $\partial \rho_{-}/\partial \xi$
and integrating them over the ES
in the normal-to-surface direction $\xi$ and using the
solutions for $w_{\pm}(x)$ up to the leading orders
[Eqs.\ (\ref{ysolplus}) and
(\ref{ysolminus})],
one arrives at the ES equations in the form of
the macroscopic boundary conditions 
\cite{strmagbr,strmagden,magsangzh,kolmagsh,magstr,magboundcond,bormot,BMRV}.
They ensure equilibrium through the equivalence of the volume and
surface (capillary) pressure
variations.
As shown in Ref.\ \cite{BMRV}, the
latter ones are proportional to the corresponding
surface tension coefficients $\sigma_{\pm}$.

For the energy surface coefficients $b_{S}^{(\pm)}$ (\ref{sigma}), one obtains
\bea\l{bsplus}
b_S^{(+)}&=& 6 \mathcal{C}_{+}
\overline{\rho} \mathcal{J}_{+}/( r_0 a),\nonumber\\
\mathcal{J}_{+}&=&\int\limits_0^1 \d w
\sqrt{w (1+\beta w)\; e_{+}[\eps(w)]},
\eea
\be\l{bsminus}
b_S^{(-)}=k^{}_S\; I^2,\qquad
k^{}_S= 6 \overline{\rho}\; \mathcal{C}_{-}\;\mathcal{J}_{-}/(r_0 a),
\ee
\bea\l{jminus}
&&\mathcal{J}_{-}=\int\limits_0^1 \d w\;
\sqrt{\frac{w\; e_{+}[\eps(w)]}{1+\beta w}}\nonumber\\
&\times&\left\{\cos(\psi) + \frac{w\sin(\psi)}{c_{\rm sym}\sqrt{1+\beta}}
\left[1+2 \widetilde{c} \gamma(w)\right]\right\}^2\nonumber\\
&\approx&\int_0^1 (1-w) \d w\;
\sqrt{\frac{w}{1+\beta w}}\;
\left\{1+\frac{2\gamma(w)}{c_{\rm sym}(1+\beta)} \right.\nonumber\\
&+&\left.\left(\frac{\gamma}{1+\beta}\right)^2\left[\frac{1}{c_{\rm sym}^2} +
6 (1+\beta)
\left(\frac{\widetilde{c}}{c_{\rm sym}} -
\frac12\right)\right]\right\}.
\eea
For $\gamma$ and $\widetilde{c}$,
see Eqs.\ (\ref{csymwt}) and (\ref{ctilde}),
respectively. Simple expressions for  the constants
$b_S^{(\pm)}$ in Eqs.\ (\ref{bsplus}) and
(\ref{bsminus}) can be easily derived  in terms of
algebraic and trigonometric functions by calculating
explicitly integrals over $w$
for the quadratic form of $e_{+}[\eps(w)]$ [Eqs. \ (\ref{Jp}) and (\ref{Jm})].
Note that in these derivations,  we neglected curvature terms
and, being of the same order,
shell corrections, which have been discarded from the very beginning.
The isovector energy terms were obtained within the ES
approximation with high accuracy up to the product of two
small quantities, $I^2$ and $(a/R)^2$.

According to the macroscopic
theory \cite{myswann69,myswnp80pr96,myswprc77,BMRV},
one may define the isovector stiffness $Q$ with respect to the difference
$R_n-R_p$
between the neutron and proton radii as a dimensionless collective variable
$\tau$,
\bea\l{Esq}
E_s^{(-)}&=&-\frac{\overline{\rho} r_0}{3} \oint
\d S\;Q \tau^2 \approx
-\frac{Q\tau^2 \mathcal{S}}{4 \pi r_0^2},\nonumber\\
  \tau&=&\left(R_n-R_p\right)/r_0\;,
\eea
where $\tau$ is the relative neutron skin.
Comparing this expression to Eq.\ (\ref{Espm})
for the isovector surface energy written through
the isovector surface-energy constant
$b_{S}^{(-)}$ [Eq.\ (\ref{bsminus})], one obtains
\be\l{stifminus}
Q=-k^{}_S\frac{I^2}{\tau^2}\;.
\ee
Defining the neutron and proton radii $R_{n,p}$ as positions of
maxima of the neutron and proton density gradients, respectively, one obtains
 the neutron skin $\tau$ (Ref.\ \cite{BMRV}),
\be\l{skin}
\tau=\frac{8 a I}{r_0 c_{\rm sym}^2}g(w_r),
\ee
where
\bea\l{fw}
g(w)&=&\frac{w^{3/2}(1+\beta w)^{5/2}}{
(1+\beta)(3w+1+4 \beta w)}\;\left\{w
(1+2\widetilde{c} \gamma)^2\right.\nonumber\\
&+&\left. 2\gamma  \left(1+\widetilde{c}
\gamma\right) \left[\widetilde{c}
w - c_{\rm sym}\left(1+
2 \widetilde{c} \gamma\right)\right]\right\}
\eea
is taken at the ES value $w_{r}$  [Eq.\ (\ref{boundcond})].
Finally taking into account Eqs.\ (\ref{stifminus}) and (\ref{bsminus}),
one arrives at
\be\l{stiffin}
Q=-\nu\; \frac{J^2}{k^{}_S}, \qquad
\nu=\frac{k_S^2 I^2}{\tau^2 J^2}=
\frac{9 \mathcal{J}_{-}^2}{16 g^2(w_r)},
\ee
where $\mathcal{J}_{-}$ and $g(w)$ are given by Eqs.\  (\ref{jminus})
and (\ref{fw}), respectively.
Note that $Q=-9J^2/4k^{}_S$ was predicted in
Refs.\ \cite{myswann69,myswnp80pr96} and
therefore for $\nu=9/4$ the first part of (\ref{stiffin}) which relates $Q$ with
the volume symmetry energy $J$ and the isovector surface-energy
constant $k^{}_S$,
is identical to that used in
Refs.\ \cite{myswann69,myswnp80pr96,myswprc77,myswiat85,vinas1,vinas2}.
However, in our derivations $\nu$ deviates from $9/4$ and it is
proportional to the function
$\mathcal{J}_{-}^2/g^2(w_r)$. This function depends
significantly on the SO interaction
parameter $\beta$ but not too much on the specific Skyrme force (see
Ref.\ \cite{BMRV} for details).

Notice that the approximate universal functions $w(x)$
[Eq.\  (\ref{ysolplus})
and Ref.\ \cite{BMRV}] and $w_{-}(x)$
[Eq.\ (\ref{ysolminus})] can be used in the leading order of the ES
approximation for calculations of the surface
energy coefficients
$b_{S}^{(\pm)}$ [Eq.\ (\ref{sigma})] and the neutron skin $\tau \propto I$
[Eq.\ (\ref{skin})]. As shown
in Ref.\ \cite{BMRV} and in Appendix B, here  only the particle
density distributions  $w(x)$  and $w_{-}(x)$ are needed
within the surface layer through their
derivatives [the lower limit of the integration
over $\xi$ in Eq.\ (\ref{sigma}) can be approximately extended to
$-\infty$ because there are no contributions from the internal volume region
in the evaluation of the main surface terms of the pressure and energy].
Therefore,  the surface symmetry-energy coefficient
$k^{}_S$ in Eqs.\ (\ref{bsminus}) and (\ref{Jm}) , the neutron skin
$\tau$ [Eq.\ (\ref{skin})] and the
isovector stiffness
$Q$ [Eq.\ (\ref{stiffin})]
can be approximated analytically in terms of functions
of definite critical combinations of the Skyrme parameters like $\beta$,
$c_{\rm sym}$, $a$, $\mathcal{C}_{-}$ and parameters of infinite nuclear matter
($b_{\rm V}$, $\overline{\rho}$, $K_{+}$), also the symmetry energy constants
$J$, $L$ and $K_{-}$. Thus, in the considered ES approximation,
they do not depend on the specific properties
of the nucleus (for instance, the neutron and proton numbers),
the curvature and the deformation of the nuclear surface.

\section{The Fermi-liquid droplet model}
\l{fldm}

For IVDR calculations,  the
FLD model based on the linearized
Landau-Vlasov equations for the isoscalar [$ \delta f^{}_{+}(\r,\p,t)$]
and isovector [$ \delta f^{}_{-}(\r,\p,t)$] distribution functions
can be used  in phase space
\cite{kolmagsh,denisov,belyaev},
\bea\l{LVeq}
\frac{\partial \delta f^{}_{\pm}}{\partial t}
&+&
    \frac{\p}{m_\pm^*}
   {\bf \nabla}_r  \left[ \delta f^{}_{\pm}\right.\nonumber\\
&+& \left.\delta \left(\e-\e^{}_F\right)
    \left(\delta V_{\pm}
    + V_{\rm ext}^{\pm}\right)\right]
    =\delta St^{}_{\pm}.
\eea
Here $\e=p^2/(2m_\pm^*)$ is the equilibrium quasiparticle energy ($p=|\p|$)
and $\e_{{}_{\! F}}=(p_F^{\pm})^2/(2m_\pm^*)$ is the Fermi energy.
The isotopic dependence of the Fermi momenta
$p_F^{\pm}=p^{}_F\left(1 \mp \Delta\right)~$
is given by a small parameter
$\Delta=2\left(1+F_{0}'\right)\;I/3~$.
The reason for having $\Delta$
is the difference between
the neutron and proton potential depths from the Coulomb interaction.
 The isotropic isoscalar $F_0$  and isovector
 $F_0'$ Landau interaction constants are related
to the isoscalar in-compressibility $K=6\e_{{}_{\! F}}(1+F_0)$
and the volume symmetry energy $J=2\e_{{}_{\! F}}(1+F_0')/3 $
constants of nuclear matter, respectively. The effective masses
$m_{+}^*=m(1+F_1/3)$ and $m_{-}^*=m(1+F_1^\prime/3)$ are determined in terms of
the nucleon mass $m$ by anisotropic Landau constants $F_1$ and
$F_1^\prime$.  Equations (\ref{LVeq}) are coupled
by  the dynamical variation
of the quasiparticles' selfconsistent interaction  $ \delta V_{\pm}$
with respect to the equilibrium
value $p^2/(2m_\pm^*)$. The   time-dependent external field
$V_{\rm ext}^{\pm} \propto \exp(-i \omega t)$
is periodic with a frequency $\omega$. For simplicity,
the collision term $\delta St_{\pm}$ is calculated
within the relaxation time $\mathcal{T}(\om)$ approximation
accounting for the retardation effects
from the energy-dependent self-energy
beyond the mean field approach,
$\mathcal{T}=4 \pi^2 \mathcal{T}_{0}/(\hbar \om)^2$
with the parameter $\mathcal{T}_{0} \propto A^{-1/3}$
[see Eq.\ (80) of Ref.\ \cite{belyaev} at zero temperature and
also Ref.\ \cite{kolmagsh}].

The solutions of Eq.\ (\ref{LVeq}) are related to the dynamic multipole
particle-density variations, $\delta \rho^{}_{\pm}(\r,t) \propto
Y_{\lambda 0}(\hat{r})$, where $Y_{\lambda 0}(\hat{r})$ are
the spherical harmonics and
$\hat{r}=\r/r$.
These solutions can be found in terms of the
superposition of plane waves over the angle
of a wave  vector
${\bf q}~$,
\bea\l{plwave}
\delta f_{\pm}(\p,\r,t)
&=&
\int \d \Om_{{\bf q}} Y_{\lambda 0}(\hat{q})\;
\delta f_{\pm}(\p,\q,\om)\nonumber\\
&\times& \exp\left[-i\left(\om t - {\bf q} \r \right)\right],
\eea
where $\delta f_{\pm}(\p,\q,\om)$ is the Fourier transform of the
distribution function.
The time-dependence (\ref{plwave}) is periodic as the external field
$V_{\rm ext}^{\pm}$ is also periodic with the same frequency
$\om=p_F^\pm s^\pm q/m_{\pm}^*$ where
$s^{+}=s$, and $s^{-}=s \left(NZ/A^2\right)^{1/2}$.
The factor $\left(NZ/A^2\right)^{1/2}$ accounts for conserving
the position of the mass center
for the isovector vibrations \cite{eisgrei}.
The sound velocity $s$ can be found from the dispersion equations
 \cite{kolmagsh}. The two solutions $s_n$ with $n=1,2$ are functions of
the Landau interaction constants and
$\omega \mathcal{T}$. Due to the symmetry interaction coupling 
the ``out-of-phase''
particle-density vibrations of the $s_1$ mode involve the ``in-phase'' mode
 $s_2$ inside of the nucleus.

For small isovector and isoscalar multipole
ES-radius vibrations of the finite neutron and proton
Fermi-liquid drops around the
spherical nuclear shape, one has
$\delta R_{\pm}(t) = R \alpha_S^{\pm}(t) Y_{\lambda 0} ({\hat r})\;$ with
 a small time-dependent amplitudes
$\alpha_S^{\pm}(t) = \alpha_S^\pm \exp(-i \omega t)$.
The macroscopic boundary conditions
(surface continuity and force-equilibrium equations) at the ES  are given by
Refs.\ \cite{BMRV,kolmagsh,belyaev}:
\bea\l{bound2}
u_{r}^{\pm}\Big|_{r=R} &=&
R \dot{\alpha}_S^{\pm} Y_{\lambda 0}({\hat r}),\nonumber\\
\delta \Pi_{rr}^{\pm}\Big|_{r=R} &=&
\alpha_S^\pm \overline{P}_S^\pm\;Y_{\lambda 0}({\hat r}).
\eea
The left hand sides of these equations
are the radial components
of the mean-velocity field ${\bf u}={\bf j}/\rho$  (${\bf j}$ is
the current density)
and the momentum flux tensor $\delta \Pi_{\nu\mu}$ defined both
through the moments of $\delta f(\r,\p,t)$ in momentum space
\cite{kolmagsh,belyaev}. The right-hand sides of Eq.\ (\ref{bound2})
are the ES velocities and capillary pressures.
These pressures are proportional to the isoscalar and isovector
surface-energy constants $b_S^{\pm}$ in Eq.\ (\ref{sigma}),
\be\l{pressuresurf}
\overline{P}_S^{\pm}=\frac23 \;
b_S^{\pm} \;
\overline{\rho}\; \mathcal{P}_{\pm}\;  A^{\mp1/3},
\ee
where $\mathcal{P}_{+}= (\lambda -1)(\lambda + 2)/2~$, $\mathcal{P}_{-}= 1~$.
The coefficients $b_S^{\pm}$ are essentially determined by the constants
$\mathcal{C}_{\pm}$ [Eq.\ (\ref{Cpm})]
of the energy density (\ref{enerdenin}) in front of its gradient density terms.
The conservation of the center of mass is taken into
account in the derivations of the second boundary conditions
(\ref{bound2}) \cite{kolmagsh,belyaev}. Therefore,
one has a dynamical equilibrium of the forces acting at the ES.

\section{Transition density and nuclear response}

The response function,
$\chi_{\pm}(\om)$,
is defined as a linear reaction to the external
single particle field $\hat{F}(\r)$
with the frequency $\omega$.
For convenience, we may consider this field in terms of a similar
superposition of plane waves (\ref{plwave}) as
$\delta f_{\pm}$ \cite{kolmagsh,belyaev}.
 In the following, we will consider the
long wave-length limit with
$\mathcal{V}_{\rm ext}^{\pm}(\r,t)=\alpha_{\rm ext}^{\pm,\om}(t) \hat{F}(\r)~$ and
$\alpha_{\rm ext}^{\pm,\om}(t)=\alpha_{\rm ext}^{\pm,\om}~e^{-i(\om+i\eta_o)t}\;$,
where $\alpha_{\rm ext}^{\pm,\om}$ is the amplitude and $\omega$ is the frequency
of the external field ($\eta_o=+0$).
In this limit, the one-body operator  $\hat{F}(\r)$ becomes the standard
multipole
operator, $\hat{F}(\r)=r^{\lambda}Y_{\lambda 0}(\hat{r})$ for $\lambda \geq 1$.
The response function $\chi_{\pm}(\om)$ is expressed through the
Fourier transform of the transition density
$\rho_{\pm}^{\om}(\r)$ as
\begin{equation}
\chi_{\pm}(\om)=
-\int {\rm d}\r\;\hat{F}(\r)\;
\rho_{\pm}^{\om}(\r)/\alpha_{\rm ext}^{\pm,\om}.
\label{chicollrho}
\end{equation}
The transition density $\rho_{\pm}^{\om}(\r)$ is obtained
through the
dynamical part of the particle density  $\delta \rho_{\pm}(\r,t)$ in a
macroscopic model
in terms of solutions $\delta f_\pm(\r,\p,t)$
of the Landau-Vlasov equations
(\ref{LVeq}) with the boundary conditions (\ref{bound2})
as the same superpositions
of plane waves  (\ref{plwave})
\cite{kolmagsh}:
$\delta \rho_{-}(\r,t) = \overline{\rho}\; \alpha^{-}_S
\rho_{-}^{\om}(x) \;Y_{10}(\hat{r})\;e^{-i \om t},$
where
\be\l{drhom}
\rho_{-}^{\om}(x)=\frac{qR}{j_1'(qR)}
\left[j^{}_1\left(\kappa\right)w(x)
+\frac{g_{{}_{\! V}}}{g_{{}_{\! S}}}\frac{\d w_{-}}{\d x}
\right],
\ee
\bea\l{gv}
g_{{}_{\! V}}&=&\int\limits_0^{w_0}
\d w
\frac{\sqrt{w(1+\beta w)}}{1-w}\kappa^3 j_{{}_{\! 1}}(\kappa),\\
\l{gs}
g_{{}_{\! S}}&=&\int\limits_0^{w_0} \d w\;
\kappa^3\left[1 + \mathcal{O}(\gamma^2(w))
\right],\\
 \kappa&=&\kappa_o\left[1+\frac{a}{R} x(w)\right],
\eea
$\kappa_o=qR$.
The first term in (\ref{drhom}), proportional
to the dimensionless isoscalar density $w(x)$ (in units of $\overline{\rho}$)
accounts for volume density vibrations [Eq.\ (\ref{ysolplus})].
The second term $\propto dw_{-}/dx$, where $w_{-}$
is a dimensionless isovector density (in units of $\overline{\rho} I$)
corresponds to the density variations from
a shift of the ES [Eq.\ (\ref{ysolminus})]. The particle number
and the center-of-mass position are conserved, and
$j_\lambda(\kappa)$ and $j_\lambda'(\kappa)$ 
are the spherical Bessel functions
and their derivatives.
The upper integration limit $w_{{}_{\! 0}}$ in Eqs.\ (\ref{gv})
and (\ref{gs}) is defined
as the root of a transcendent equation $x(w_{0})+R/a=0$.
As shown in 
Appendix A, the SO and $L$ dependent density $w_{-}(x)$
is of the same order as $w(x)$.
The dependencies of  $w_{-}(x)$ on
different Skyrme force parameters, mostly the isovector gradient-term constant
$\mathcal{C}_{-}$, the SO parameter $\beta$, and the
derivative of the volume symmetry energy $L$ are
the main reasons for the different values of the neutron skin.

With the help of the
boundary conditions (\ref{bound2}), one can derive the response
function (\ref{chicollrho}) \cite{kolmagsh},
\be\l{respfuni}
\chi^{}_{\lambda}(\om)=\sum_n \chi_{\lambda}^{(n)}(\omega)
=
\sum_n \mathcal{A}_{\lambda}^{(n)}(\kappa_o)/ \mathcal{D}_{\lambda}^{(n)}
\left(\om - i\frac{\Gamma}{2}\right),
\ee
with $\om=p^{}_Fs_n \kappa_o \left(NZ/A^2\right)^{1/2}/(m^*R)~$
($m_{-}^*\approx m_{+}^*=m^*$).
This response function describes two modes, the main ($n=1$)
IVDR and its satellite  ($n=2$)
as related to the out-of-phase $s_1$ and in-phase $s_2$ sound velocities
which are excited in the nuclear volume,
respectively. We assume here that the ``main''
peak exhausts mostly
the energy weighted sum rule (EWSR) and the ``satellite'' corresponds
to a much smaller part of the EWSR as proportional to the
asymmetry parameter, $I \ll 1$. This two-peak structure is from
the coupling of the
isovector and isoscalar density-volume vibrations because of
 the neutron and proton
quasiparticle interaction $\delta V_{\pm}$ in Eq.\ (\ref{LVeq}).
Therefore, one takes into account
an admixture of the isoscalar mode to the isovector IVDR excitation.
The wave numbers $q=\kappa_o/R$ of the lowest poles ($n=1,2$)
 in the response function (\ref{respfuni})
are determined by the secular equation,
\be\l{seculeq}
\mathcal{D}_{\lambda}^{(n)} \equiv  j_{\lambda}'(\kappa_o)
-
\frac{3 \e_{{}_{\! F}}\kappa_oc_1^{(n)}}{2 b_S^{-} A^{1/3}}
\left[j_{\lambda}(\kappa_o) + c_2^{(n)}j_{\lambda}''(\kappa_o)\right]
=0.
\ee
The width of an IVDR peak $\Gamma$ in (\ref{respfuni})
corresponds to an imaginary part
of the pole having its origin in the  collision term
$\delta St_{\pm}$
of the Landau-Vlasov equation. At this pole, for the relaxation time
one has
\be\l{relaxtimen}
\mathcal{T}_n=4 \pi^2 \mathcal{T}_{0}/(\hbar \om_n)^2
\ee
with an $A$-dependent constant, $\mathcal{T}_0 \propto A^{-1/3}$.
For the amplitudes one has $\mathcal{A}_{\lambda}^{(n)} \propto \Delta^{n-1}$.
The complete expressions for the amplitudes $\mathcal{A}_{\lambda}^{(n)}$ and the
constants $c_i^{(n)}$ are given in Refs.\ \cite{kolmagsh,belyaev}.
Assuming a small value of $\Delta$, one may
call the $n=2$ mode a ``satellite''  to the ``main'' $n=1$ peak.
On the other hand, other factors such as a  collisional relaxation  time,
 the surface symmetry-energy constant $b_{S}^{-}$, and the particle
number $A$ lead sometimes to a re-distribution
of the EWSR values among these two IVDR peaks. The slope $L$ dependence
of the transition densities $\rho_{-}^{\om}(x)$ [Eq.\ (\ref{drhom})] and
the strength of the response function,
\be\l{strength}
S(\omega)=\Im \chi^{}_{\lambda}(\om)/\pi
\ee
[Eq.\ (\ref{respfuni})] has its origin in the symmetry-energy
coefficient $b_S^{(-)}$ [Eqs.\ (\ref{bsminus}), (\ref{jminus}),
(\ref{ctilde}), (\ref{epsilonplus}) and (\ref{symen})].  Thus,
one may evaluate the EWSR sum rule contribution of the $n$th peak by
integration over the region around the peak energy
$E_n=\hbar \om_n$,
\be\l{sumruleewsr}
S_n^{(1)} =\hbar^2\int  \d \omega\; \omega\;S_n(\omega).
\ee

In accordance with the time-dependent HF approaches based on the Skyrme forces,
(see, for instance,
\cite{nester1,nester2,Endressplitting}), we may expect that 
the energies of the 
satellite resonances in the IVDR and ISDR channels 
can be close.  Therefore, we may calculate separately the
neutron, $\rho_n^{\om}(x)$, and proton, $\rho_p^{\om}(x)$, transition
densities for the satellite by calculating the isovector and isoscalar
transition densities at the same energy $E_2$
and in the same units as $\rho_{\pm}$,
\be\l{rhonp}
\rho^{\om}_{n}(x)=
\frac{\rho_{+}^{\om}(x) + \rho_{-}^{\om}(x)}{2},\quad
\rho^{\om}_{p}(x)=
\frac{\rho_{+}^{\om}(x) - \rho_{-}^{\om}(x)}{2}.
\ee

\section{Discussion of the results}

In Table II we show the isovector surface energy coefficient $k^{}_S$ 
[Eq.\ (\ref{bsminus})],
the stiffness parameter $Q$ [Eq.\ (\ref{stiffin})], its constant $\nu$ and
the neutron skin $\tau$ [Eq.\ (\ref{skin})] 
for many more Skyrme forces
than discussed in \cite{BMRPhysScr2015}. They are obtained within the ES
approximation with the quadratic expansion for $e_{+}[\eps(w)]$ and neglecting
the $I^2$ slope corrections, for several Skyrme forces
\cite{chaban,reinhardSV} whose parameters are presented in Table I.
Also shown are the quantities $k_{S 0}$, $\nu_0$, $Q_0$ and $\tau_0$
neglecting the slope corrections ($L=0, K_{-}=0$). This is in addition to
results of Ref.\ \cite{BMRV} where another important dependence
on the SO interaction measured by $\beta$ was presented. 
In contrast to a fairly good agreement for the analytical isoscalar
surface-energy constant $b_S^{(+)}$ (\ref{bsplus}) as shown in Ref.\
\cite{BMRV} and references cited therein,
the isovector energy coefficient $k^{}_S$ is more sensitive to the choice
of the  Skyrme forces than the isoscalar one $b_S^{(+)}$
[Eq.\ (\ref{bsplus}) and Ref.\ \cite{magsangzh}].
The modulus of  $k^{}_S$  is significantly larger for most of
the Skyrme forces SLy...
\cite{chaban} and SV... \cite{reinhardSV} than for the other ones. However,
the $L$ dependence of $k^{}_S$ is somewhat small in these forces
(cf. the first two rows of Table II)
as it should be for  a small parameter $\eps$ of the symmetry-energy density
expansion (\ref{symen}). For SLy and SV forces,
the stiffnesses $Q$ are correspondingly significantly smaller in absolute value
being closer to the well-known empirical
values $Q \approx 30-35$ MeV \cite{myswprc77,myswnp80pr96,myswiat85}
obtained by Swiatecki and collaborators.
Note that the isovector stiffness $Q$ is even much more sensitive
to the parametrization of the Skyrme force and
to the slope parameter $L$ than the
constants $k^{}_S$. In Ref.\ \cite{BMRV}, we studied the
hydrodynamical results for $Q$
as compared to the FLD model for the averaged properties of  the giant
IVDR (IVGDR) at zero slope $L=0$.
The IVDR structure in terms of the two (main and satellite)
peaks was discussed 
earlier in Refs.\ \cite{BMRPhysScr2014} 
at $L=0$ in some
magic nuclei  with a large neutron excess within the semiclassical FLD model
based on the effective surface approach.
For the comparison with experimental data and other theoretical results
we present in Table II (row 9 and 11) a small $L$ dependence of the
IVGDR energy parameter
$D=E_{\rm IVGDR}A^{1/3}$, where
$E_{\rm IVGDR}=\left[E_1 S_1 +E_2 S_2\right]/\left[
S_1 + S_2\right]~$
is the IVGDR energy for the 
isotope $^{132}$Sn
[$~S_n=S(\om_n)~$, see also
Eq.\ (\ref{strength}) for the definition of the strength $S(\om)$].
A more precise reproduction of the $A$-dependence of the IVGDR energy
parameter $D$
for finite values of $L$ (see the last three rows for several isotopes)
might determine more consistent values of $Q$, but, at present, it seems
to be beyond the accuracy of both the hydrodynamical and the FLD models.
The IVGDR energies obtained  
by the semiclassical Landau-Vlasov equation
(\ref{LVeq}) with the macroscopic boundary conditions (\ref{bound2})
of the FLD model
(Ref.\ \cite{BMRV}) are also basically insensitive
to the isovector surface energy constant $k_S$ 
\cite{BMV,BMRV,BMRPhysScr2014,BMRPhysScr2015}.
They are in a good agreement with the experimental data, and do not depend much
on the Skyrme forces even if we take into account the slope symmetry-energy
parameter $L$ (last three rows in Table II).

More realistic self-consistent  HF calculations taking into account the Coulomb
interaction, the surface curvature, and quantum shell effects have led to
larger
values of $Q\approx 30-80$ MeV \cite{brguehak,vinas2}.
For larger $Q$ (see Table II)
the fundamental parameter $(9 J /4Q) A^{-1/3}$ of the LDM expansion
in Ref.\ \cite{myswann69} is really small for $A \simg 40$, and therefore,
the results obtained using  the leptodermous expansion are better
justified.

An investigation within the approach presented in
Sec.\ V shows that the IVDR strength
is split into a main peak which
exhausts an essential part of the EWSR independent of the model
and a satellite peak
with a much smaller contribution to this quantity.  Focusing on a much more
sensitive  $k_S$ dependence of the  
IVDR satellite resonances,
one may take now into account the slope $L$ dependence of the
symmetry-energy density per particle (\ref{symen})
(Refs.\ \cite{kievPygmy,larsenprc2013,nester1,nester2}
and \cite{BMRPhysScr2014}).
The total IVDR strength function being
the sum of the  ``out-of-phase'' $n=1$ and
``in-phase'' $n=2$ modes for the isovector- and
isoscalarlike
volume particle density vibrations, respectively
(solid lines in Figs.\ \ref{fig3} and \ref{fig4} for
the zero $L$ and dotted and dashed lines for the finite $L$)
has a somewhat remarkable shape asymmetry \cite{BMRPhysScr2014,BMRPhysScr2015}.
For SLy5$^*$ (Fig.\ \ref{fig3}) and  for SVsym32 (Fig.\ \ref{fig4})
one has the ``in-phase'' satellite to the right of the
main ``out-of-phase'' peak, 
cf.\ with the traditional PDR to the left
of the main one. 
An enhancement to the left of the main peak
for SLy5* is from increasing the
``out-of-phase'' strength (red solid and magenta rare 
dotted curves, Fig.\ \ref{fig3})
at small energies
because of the appearance of a peak at the energy about a few
MeV, in contrast to the SVsym32 case. The semiclassical FLD model
calculations at the lowest $\hbar$ order should be improved here, for
instance by taking into account the quantum effects like shell
corrections within more general periodic-orbit theory \cite{belyaev,BM}.
In the nucleus $^{132}$Sn the IVDR energies of the two peaks
 do not change much with $L$ in both cases: $E_1=17$ MeV, $E_2=20$ MeV
for SLy5$^*$ (Fig.\ \ref{fig3}) and $E_1=15$ MeV, $E_2=18$ MeV for SVsym32
(Fig.\ \ref{fig4}). 
We find only an essential re-distribution of the EWSR contributions
(normalized to 100\% for the EWSR sum of the main 
and satellite peaks) [Eq.\ (\ref{sumruleewsr}) for $S_n^{(1)}$].
This is from a significant enhancement of
the main ``out-of-phase'' peak
with increasing $L$,  $S_1^{(1)}=89$\% and $S_2^{(1)}=11$\% for SLy5$^*$
(Fig.\ \ref{fig3}) and more pronounced EWSR
distribution $S_1^{(1)}=76$\% and $S_2^{(1)}=24$\% for SVsym32
(Fig.\ \ref{fig4}) [cf.\ with the corresponding $L=0$ results:
$S_1^{(1)}=88$\% and $S_2^{(1)}=12$\% for SLy5* and
$S_1^{(1)}=73$\% and $S_2^{(1)}=27$\% for SVsym32].

Figures \ref{fig5} and \ref{fig6} 
show more systematic study for several isotopes and for the chain of the
Sn isotopes, respectively. In Fig.\ \ref{fig6}, we compare the results
of our calculations with the experimental data. The latter
were obtained by fitting the
experimental strength curve for a given almost spherical Sn isotope 
by the two Lorenzian oscillator strength functions as described in Refs.\ 
\cite{kolmagsh,belyaev}. It is always possible in the case of the asymmetric 
shapes of the strength curves with usual enhancement on the right of the main 
peak, even in the case when the satellite cannot be distiguished transparently 
well from
the main peak in almost spherical nuclei (unlike the clear shoulders
for the IVDRs in deformed ones). 
Each of these functions has three fitting parameters
such as the inertia, stiffness and width of the peak. 
We found somewhat good agreement
of our ETF ES results with these experimental data for the energies, ratio of
the strengths at the satellite to the main modes, and the EWSR contributions.

More precise $L$-dependent calculations
change essentially the IVDR strength distribution
for the SV forces  because of the smaller $c_{\rm sym}$ value as compared
to other Skyrme interactions (see Table I). For $^{208}$Pb one obtains
$E_1=15$ MeV, $S_1^{(1)}=91$\% for the main peak 
and $E_2=17$ MeV, $S_2^{(1)}=9$\% the satellite for SLy5$^*$; and 
$E_1=13$ MeV, $S_1^{(1)}=83$\% for the main peak 
and $E_2=16$ MeV, $S_2^{(1)}=17$\% the satellite for SVsym32 forces. 
These calculations
are qualitatively in agreement with the experimental
results: $E_1=13$ MeV, $S_1^{(1)}=98$\% for the main peak 
and $E_2=17$ MeV, $S_2^{(1)}=2$\% the satellite. 
Discrepancies might be related
to the strong shell effects in this stable double magic nucleus 
which are neglected in the ETF ES approach.

Decreasing the relaxation time $\mathcal{T}$ by
a factor of about 1.5 almost does not change the IVDR strength structure.
However,
we found a strong dependence on the relaxation time $\mathcal{T}$ in a wider
region of $\mathcal{T}$ values. The ``in-phase'' strength component with
a wide maximum does not depend much on the Skyrme force
\cite{chaban,reinhardSV,pastore}, the slope parameter
$L$, and the relaxation time $\mathcal{T}$.
We found also a regular change of the
IVDR strength for different double magic
isotopes (Fig.\ \ref{fig5}). In addition to a big
change for the energy (mainly because of $E_1$) and the strength
[$S_1(\omega$)], one also obtains more asymmetry
for $^{68}$Ni than for the other isotopes.
Calculations for nuclei with different mass $A$
were performed with the
relaxation time  $\mathcal{T}$ [Eq.\ (\ref{relaxtimen})]
where
$\mathcal{T}_{0}=\mathcal{T}_{0\rm Pb} (208/A)^{1/3}$ with the parameter
$\mathcal{T}_{0\rm Pb}=300 $ MeV$^2\cdot$s derived from the IVGDR width
of $^{208}$Pb, in agreement with experimental data for the averaged
$A$ dependence of the IVGDR widths ($\propto A^{-2/3}$).
 In this way the IVDR relaxation time $\mathcal{T}_n$  becomes
larger with increasing
$A$ as $A^{1/3}$, and at the same time, the height of peaks  decreases.
The $L$ corrections are also changing much in the same scale of all three
nuclei.

The essential parameter of the Skyrme HF approach leading to the
significant differences in the $k_S$ and $Q$ values is the constant
$\mathcal{C}_{-}$
[Eq.\ (\ref{enerdenin}) and Table I].
Indeed, $\mathcal{C}_{-}$ is the key quantity in the expression for $Q$
[Eq.\ (\ref{stiffin})]
and the isovector surface-energy constant $k^{}_S$ [or $b_S^{(-)}$,
Eq.\ (\ref{bsminus})],
because $Q \propto 1/k_S \propto 1/\mathcal{C}_{-}$
and $k^{}_S\propto \mathcal{C}_{-}$ \cite{BMRV}. Concerning $k^{}_{S}$
and the IVDR strength structure,  this is even more
important than the $L$ dependence although the latter changes significantly the
isovector stiffness $Q$ and the neutron skin $\tau$.
As seen in Table I,  the constant $\mathcal{C}_{-}$
is very different in absolute value and in sign for different Skyrme
forces whereas $\mathcal{C}_{+}$ is almost constant (Table I).
The isoscalar energy density constant $ b_S^{(+)}$
is proportional to $C_{+}$ [Eq.\ (\ref{bsplus})],
in contrast to the isovector one.
All Skyrme parameters are fitted to the well-known experimental
value  $b_S^{(+)} =17-19$ MeV  while
there are so far no clear experiments which would determine $k^{}_S$
well enough because the mean energies of the IVGDR (main peaks) do not
depend very much on $k^{}_S$
for different Skyrme forces (the last three rows of Table II).
Perhaps, the low-lying isovector collective states are more sensitive but
at the present time there is no careful systematic study of their
 $k^{}_S$ dependence.
Another reason for so different $k_S$ and $Q$ values might be because of
 difficulties in deducing $k^{}_S$ directly from the HF calculations
because of the curvature and quantum effects.
In this respect, the semi-infinite Fermi system with a hard plane wall
might be more adequate for the comparison of the
HF theory and  the ETF effective surface approach.
We have also to go far away from the nuclear stability line to
subtract uniquely the coefficient $k^{}_S$ in the dependence of
$b_S^{(-)} \propto I^2=(N-Z)^2/A^2$, according to Eq.\ (\ref{bsminus}).
For exotic nuclei one has more problems to  derive $k^{}_S$  from the
experimental data with enough precision.
Note that, for studying the IVDR structure, the quantity $k^{}_S$
is more fundamental
 than the isovector stiffness $Q$  because of the direct relation to the
tension coefficient $\sigma_{-}$ of the isovector capillary pressure.
Therefore, it is simpler to analyze
the experimental data for the IVGDR within the macroscopic HD or FLD models
in terms of the constant $k^{}_S$. The quantity $Q$ involves also the
ES approximation for
the description of the nuclear edge through the neutron skin $\tau$ in
 Eq.\ (\ref{stifminus}). The $L$ dependence of the neutron skin $\tau$
is essential but not so dramatic in the case of SLy and SV
forces (Table II),
except for the SVmas08 forces with the effective mass
0.8. The precision of such a description depends more on the specific
nuclear models \cite{vinas1,vinas2,vinas5}.
On the other hand, the neutron skin thickness $\tau$, like the stiffness $Q$,
is interesting in many aspects for an investigation of exotic nuclei,
in particular, in nuclear astrophysics.

We emphasize that for specific Skyrme forces there exists an
abnormal behavior of the
isovector surface constants $k^{}_S$ and $Q$. It is related to the
fundamental constant $\mathcal{C}_{-}$ of
the energy density (\ref{enerdenin}) but not to
the derivative symmetry-energy density corrections. For the parameter
set T6  ($\mathcal{C}_{-}=0$) one finds $k^{}_S=0$ (Ref.\ \cite{BMRV}).
Therefore,
according to Eq.\ (\ref{stiffin}), the value of $Q$ diverges
($\nu$ is almost independent from $\mathcal{C}_{-}$ for SLy and SV forces;
 Table II and Refs.\
\cite{BMRV,BMRPhysScr2014,BMRPhysScr2015}).
The isovector gradient terms which are important for
the consistent derivations within the ES approach are also not included
($\mathcal{C}_{-}=0$) into the symmetry energy density in 
Refs.\ \cite{danielewicz1,danielewicz2,danielewicz3}.
 In relativistic investigations \cite{vretenar1,vretenar2} of
the structure
of the IVGR distributions, the dependence of these quantities
on the derivative terms has not
been investigated so far. It therefore remains an interesting
task for the future to
apply similar semiclassical methods such as the  ES approximation
 used here also in
relativistic models.
Moreover, for RATP \cite{chaban} and SV \cite{reinhardSV}
(like for SkI) Skyrme forces, the isovector stiffness $Q$ is even negative
as $\mathcal{C}_{-}>0$ ($k^{}_S>0$) in contrast to other Skyrme forces.
This would lead to an instability of the vibration of the neutron skin.

Table II shows also the coefficients $\nu$ of Eq.\ (\ref{stiffin})
for the isovector stiffness $Q$.
They are almost constant for all SLy and SV
Skyrme forces, in contrast to other forces \cite{BMRV}.
However, these constants $\nu$, being sensitive to the SO
($\beta$) dependence through Eqs.\  (\ref{fw}),  (\ref{skin}) and
(\ref{jminus}), change also with $L$ (Table II).
As compared to 9/4 suggested in Ref.\ \cite{myswann69}, they are
significantly smaller
in magnitude for most of the Skyrme forces.

In Fig.\ \ref{fig7} we show, in the case of the Skyrme forces SLy5*
and SVsym32, the transition densities $\rho_{\mp}^\omega(x)$ of
Eq.\ (\ref{drhom}) for the ``out-''of-phase (-) and the ``in-''phase (+)
modes of the volume vibrations at the excitation
energy $E_2$ of the satellite. 
The transition densities were not presented
in our preliminary publication \cite{BMRPhysScr2015}.
These are the key quantities for the calculation of the IVDR strengths,
according to Eq.\ (\ref{chicollrho}).
The $L$ dependence is somewhat small, slightly notable mostly near the ES
($|x|\siml 1$). From Fig.\ \ref{fig8}, one finds a remarkable neutron vs
proton excess  near the nuclear edge for the same forces, which is however,
very slightly depending on the slope parameter $L$. A small dependence
of the transition densities on $L$ comes through
the symmetry-energy constant $k^{}_S$ which is almost the same
in modulus for these forces. We did not find a dramatic change of
the transition densities with the sign of $k^{}_S$.
Therefore, there is a weak sensitivity of the transition densities on $L$
through the energy $E_2$. We would have expected a stronger influence of
the sign of $k_S^{}$ on the vibrations of the neutron skin rather than on the
IVDR.
This different sign leads to the opposite, stable and unstable,
neutron skin vibrations.  One observes also other differences
between the upper (SLy5*) and the
lower (SVsym32) panels in both figures:
We find a redistribution of the surface-to-volume
contributions of the transition densities for these two modes.
Again, as in Figs.\ \ref{fig9} and \ref{fig10}, one finds
a considerable change of
the neutron-proton transition densities for the different isotopes for SLy5*
and SVsym32.

The last figure shows theoretical (Fig.\ \ref{fig11})
 evaluations of the neutron skin.
Figure \ref{fig11} shows the absolute values of
the skin obtained from $\tau/I$ multiplying the mean-square evaluations of
the nuclear radii by the factor $\sqrt{3/5}$ for an easy comparison with  
experimental data given in \cite{vinas5}. 
For $^{208}$Pb, one finds that the experimental values 
$\Delta r_{np}^{exp}=0.12-0.14$ fm  in Ref.\ \cite{vinas5} 
(0.156$_{-0.021}^{+0.025}$ fm \cite{rcnp}) are in good agreement with our
calculations $\Delta r_{np}^{theor}\approx 0.10-0.13$ fm
within the ES approximation
(the limits show values from SLy5* to SVsym32). For the isotope
$^{124}$Sn one obtains $\Delta r_{np}^{theor}\approx 0.09-0.12$ fm, also 
in good agreement with experimental results. For the isotope
$^{132}$Sn, we predict the value
$\Delta r_{np}^{theor}\approx 0.11-0.15$ fm.
Similarly, for $^{60}$Ni and $^{68}$Ni, one finds
 $\Delta r_{np}^{theor}\approx 0.03-0.04$ (like in Ref.\ \cite{vinas5})
and $0.08-0.11$ fm, respectively.

\section{Conclusions}

 The slope parameter $L$ was taken into account in the leading ES
approximation to derive simple analytical expressions  for
the isovector
particle densities and energies. These expressions were used for calculations of
the surface symmetry energy, the neutron skin thickness,
and the isovector stiffness coefficients as functions of
$L$. For the derivation of the surface symmetry energy and its dependence
on the particle density we have to include main higher order terms
in the parameter $a/R$.
These terms depend on the well-known parameters of
the Skyrme forces. Results for
the isovector surface-energy constant $k_S$,  the neutron skin
thickness $\tau$ and the stiffness $Q$
depend in a sensitive  way on the parameters of the
Skyrme functional (especially on the
parameter $\mathcal{C}_{-}$) in the gradient terms of the density in the
surface symmetry energy [see Eq.\ (\ref{enerdenin})].
The isovector constants $k_S$, $\tau$ and $Q$ depend also
essentially on the slope parameter $L$, 
in addition to the SO interaction constant $\beta$.
For all Skyrme forces, the isovector stiffness constants
$Q$ are significantly larger than those obtained in earlier investigations.
However, taking into account their $L$-dependence they come closer
to the empirical data. It influences 
more on the isovector stiffness $Q$ and on the neutron skin $\tau$,
than on the surface symmetry-energy constant $k^{}_S$.
The mean IVGDR energies and sum rules calculated in the
macroscopic models like the FLD  model \cite{kolmagsh,BMV}
in Table II are in fairly good agreement with the
experimental data for most of the $k_S $ values. As compared with
the experimental data and other recent theoretical works, we found a
somewhat reasonable two-peak
structure of the IVDR strength within the FLD model.
According to our results for  the neutron and proton transition densities
[Figs.\ \ref{fig8}-\ref{fig10}],
we may interpret semiclassically the IVDR satellites as
some kind of pygmy resonances, 
in addition to the traditional studies
\cite{vretenar1,vretenar2,ponomarev,kievPygmy,larsenprc2013,nester1,nester2,adrich,wieland,reviewpygmy1,reviewpygmy2}.
Their energies, sum rules and n-p transition densities
obtained analytically within the semiclassical FLD approximation
are sensitive to the surface symmetry-energy constant
$k_{{}_{\! S}}$ and the slope parameter $L$. Therefore, their comparison with the
experimental
data can be used for the evaluation of $k_{{}_{\! S}}$ and $L$.
It seems helpful to describe them in terms of only few critical
parameters, like $k^{}_S$ and $L$.

For further perspectives, it would be worthwhile to
apply our results to calculations of the 
IVDR strength structure within the FLD model \cite{kolmagsh} 
in a more systematic way.
In this respect it is also interesting that the low-lying collective
isovector states are expected to be even
more sensitive to the values of $k_S$ within the 
periodic orbit theory \cite{gzhmagfed,blmagyas,BM}.
More general problems of classical and quantum chaos
in terms of the level statistics and Poincare and Lyapunov exponents
(Ref.\ \cite{BMstatlev2012} and references therein) might lead to progress
in studying the fundamental
properties of collective dynamics like nuclear fission within the 
Swiatecki-Strutinsky
macroscopic-microscopic model. Our approach is helpful also
for further study of the effects in the surface symmetry energy because it
gives analytical
universal expressions for the  constants $k^{}_S$, $\tau$ and $Q$
as functions of the slope parameter $L$ which do not depend on
specific properties of nuclei as they are directly connected with
a few critical parameters of the Skyrme interaction without 
any fitting.

\bs
\centerline{{\bf Acknowledgements}}
\ms

The authors thank
V.I.\ Abrosimov, K.\ Arita, V.Yu.\ Denisov, V.M.\ Kolomietz, M.\ Kowal,
M.\ Matsuo, K.\ Matsuyanagi, J.\ Meyer, V.O.\ Nesterenko,
M.\ Pearson, V.A.\ Plujko, P.-G.\ Reinhard, A.I.\ Sanzhur, J.\ Skalski,
and X.\ Vinas for many useful discussions. One of us (A.G.M.) is
also very grateful for the warm hospitality during working visits at the
National Centre for Nuclear Research in Warsaw, Poland, and also
for financial support from the Japanese
Society of Promotion of Sciences, Grant No. S-14130,
during his stay at the Nagoya Institute of Technology.
This work was partially supported by the Deutsche
Forschungsgemeinschaft Cluster of Excellence 'Origin and
Structure of the Universe' (www.universe-cluster.de).

\appendix \setcounter{equation}{0}
\renewcommand{\theequation}{A\arabic{equation}}

\begin{center}
\textbf{Appendix A: SOLUTIONS OF THE ISOVECTOR LAGRANGE EQUATION}
\label{appA}
\end{center}

\begin{figure*}
\begin{center}
\includegraphics[width=1.0\columnwidth]{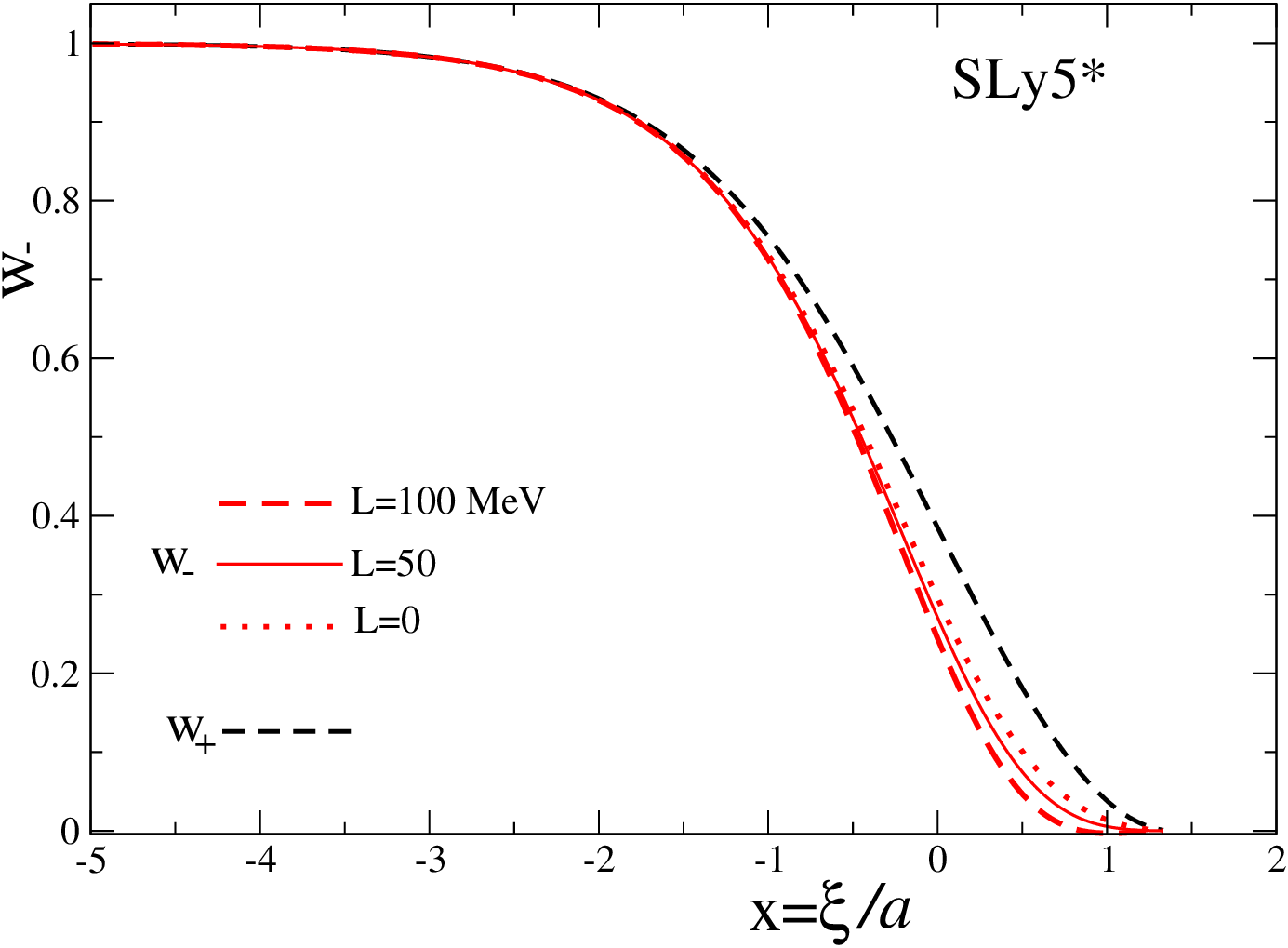}
\end{center}

\vspace{-0.3cm}
\caption{(Color online) Isovector $w_{-}$
(\ref{ysolminus}) (with the relevant value of $L$
Refs.\ \cite{vinas2,reinhardSV,jmeyer})
and without ($L=0$) derivative $L$ constant, and isoscalar $w=w_{+}$
(see \cite{BMRV}) particle densities are shown
vs $x=\xi/a $
for the Skyrme force SLy5$^*$ ($x\approx (r-R)/a$ for small nuclear
deformations \cite{pastore,BMRPhysScr2014,BMRPhysScr2015}).
}
\label{fig1}
\end{figure*}

\begin{figure*}
\begin{center}
\includegraphics[width=0.6\textwidth]{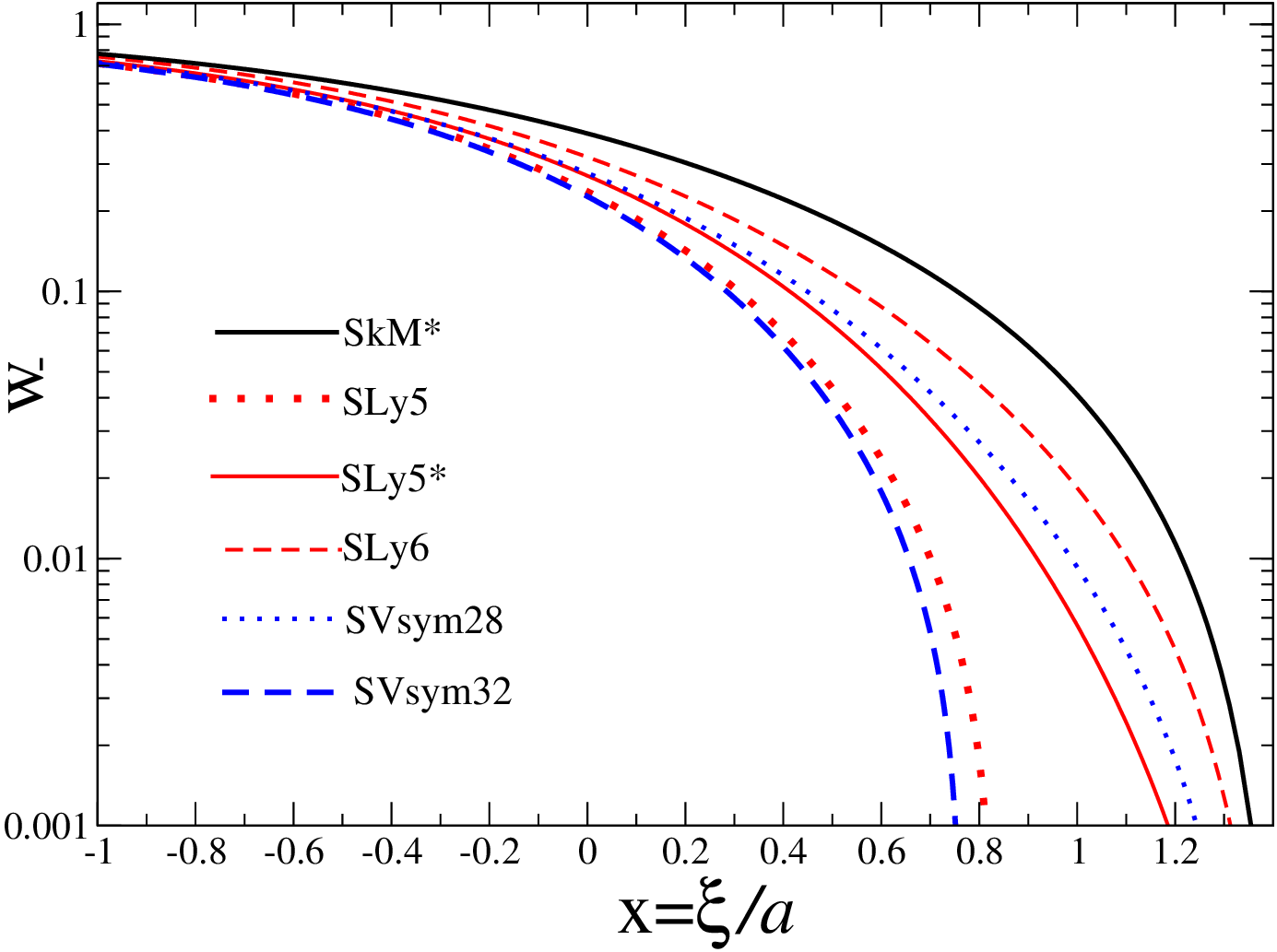}
\end{center}

\vspace{0.2cm}
\caption{(Color online)
Isovector density
$w_{-}(x)$ (\ref{ysolminus}) (in the logarithmic scale)  as function of $x$
within the quadratic approximation to $e_{+}[\eps(w)]$
for several Skyrme forces \cite{chaban,reinhardSV,vinas2,pastore,jmeyer}.
}
\label{fig2}
\end{figure*}

\begin{figure*}
\begin{center}
\includegraphics[width=0.6\textwidth]{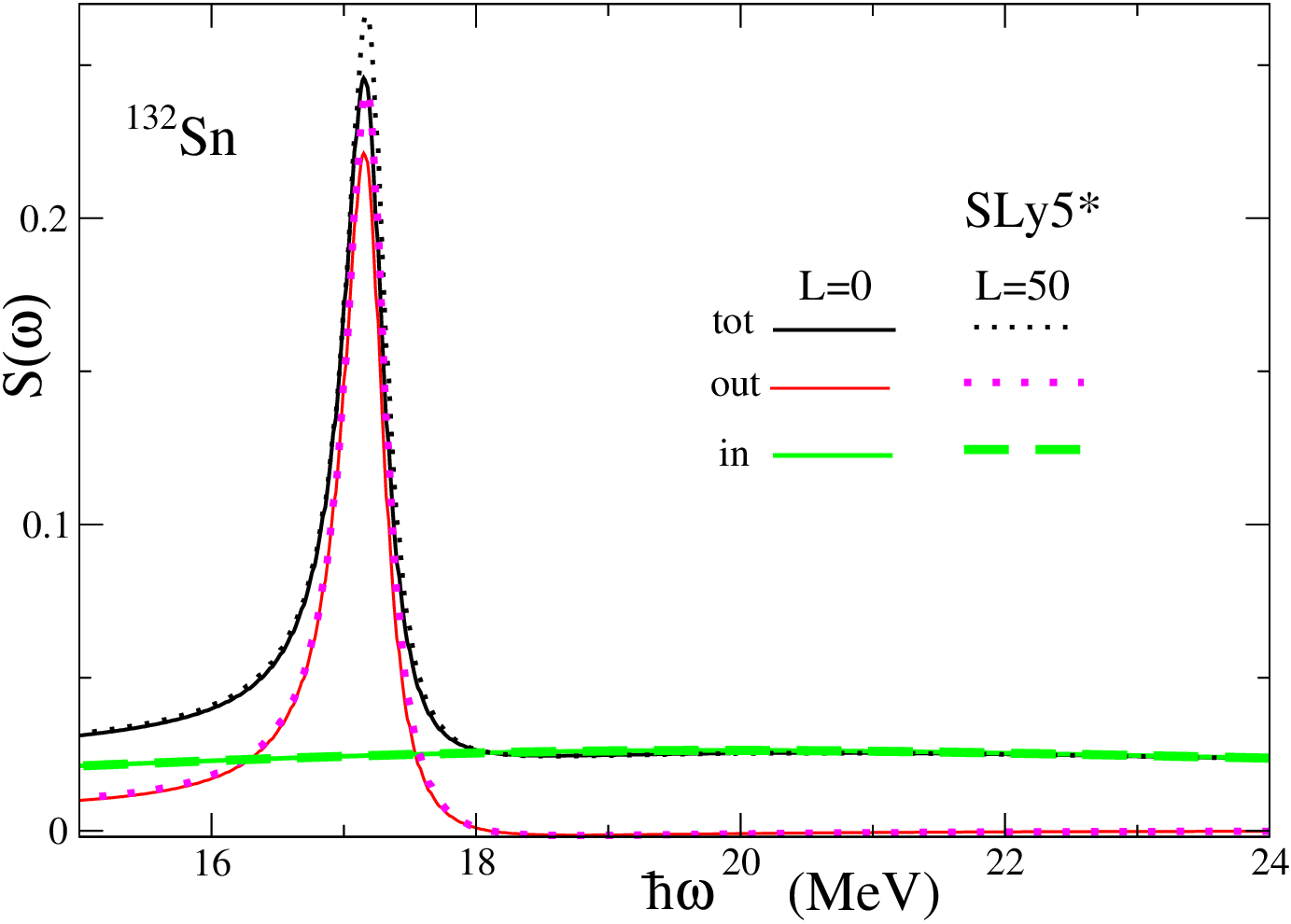}
\end{center}

\vspace{-0.1cm}
\caption{(Color online) IVDR strength functions
$S(\omega)$
vs the excitation
energy $\hbar \omega$ are shown for vibrations of the nucleus
$^{132}$Sn for the Skyrme force SLy5$^*$ by dots and dashed lines at $L=50$ MeV
and solid lines for $L=0$;
red or magenta (``out-of-phase''), and  green (``in-phase'')
curves show separately the main and satellite excitation modes, respectively
(section 4 and 5);
the collision relaxation time
$\mathcal{T}=4.3 \cdot 10^{-21}$ s in agreement with the IVGDR  widths
\cite{belyaev}.}
\label{fig3}
\end{figure*}

\begin{figure*}
\begin{center}
\includegraphics[width=0.6\textwidth]{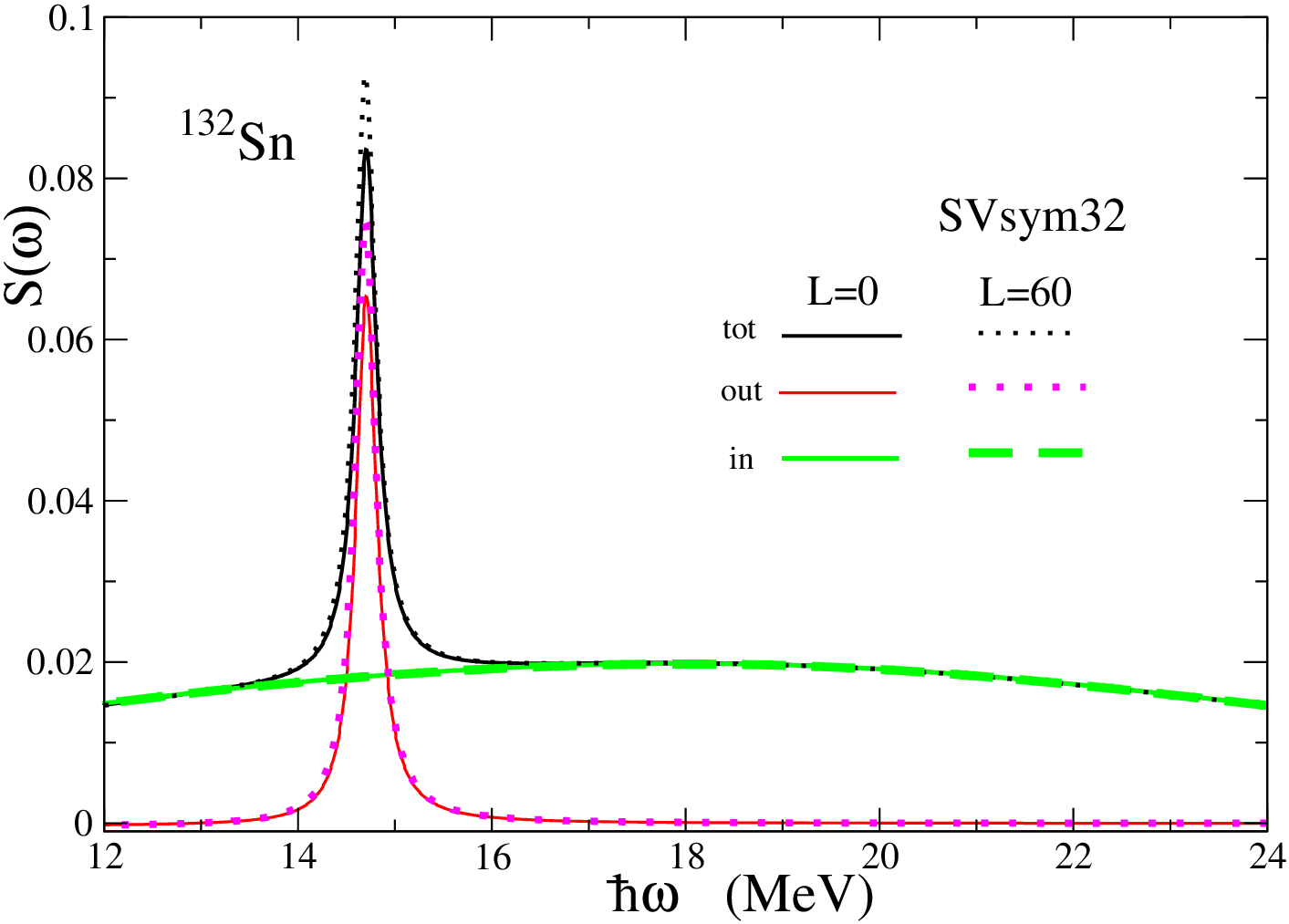}
\end{center}

\vspace*{0.4cm}
\caption{(Color online)
The same total and different modes (main and satellite) strengths
as in Fig.\ \ref{fig3} are shown
for different $L=0$ and $60$ MeV for the Skyrme force SVsym32.
}
\label{fig4}
\end{figure*}

\begin{figure*}
\begin{center}
\includegraphics[width=0.80\textwidth]{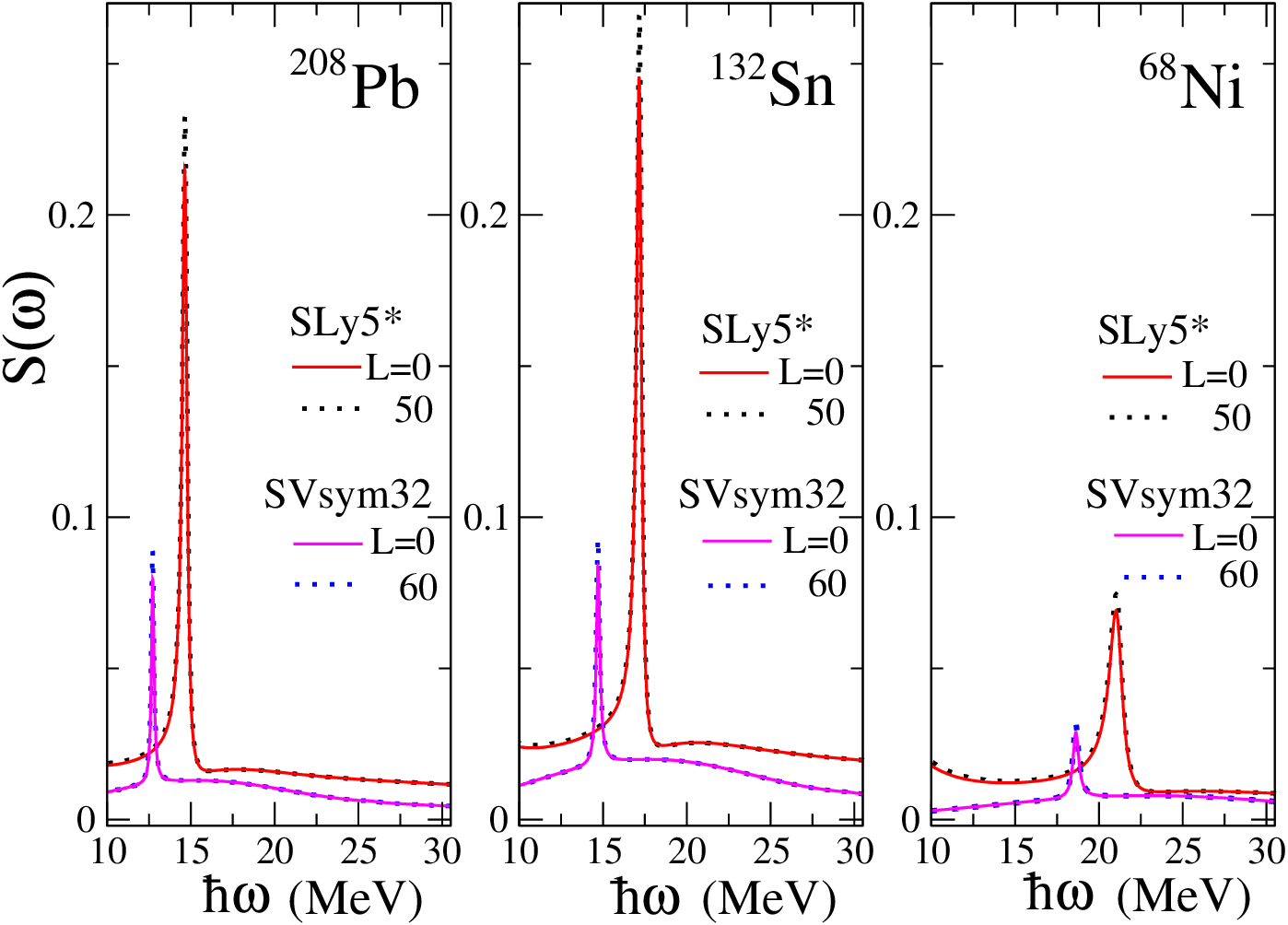}
\end{center}

\vspace{0.5cm}
\caption{(Color online)
The total IVDR strength functions $S(\omega)$ vs the excitation energy
$\hbar \omega$ (in MeV) for different double magic nuclei for SLy5*
and SVsym32 forces; a slight dependence on the
slope parameter $L$ (in MeV) as compared to the $L=0$ case at the 
main peaks is shown.}
\label{fig5}
\end{figure*}

\begin{figure*}
\begin{center}
\includegraphics[height=14.0cm,width=14.0cm,angle=0]{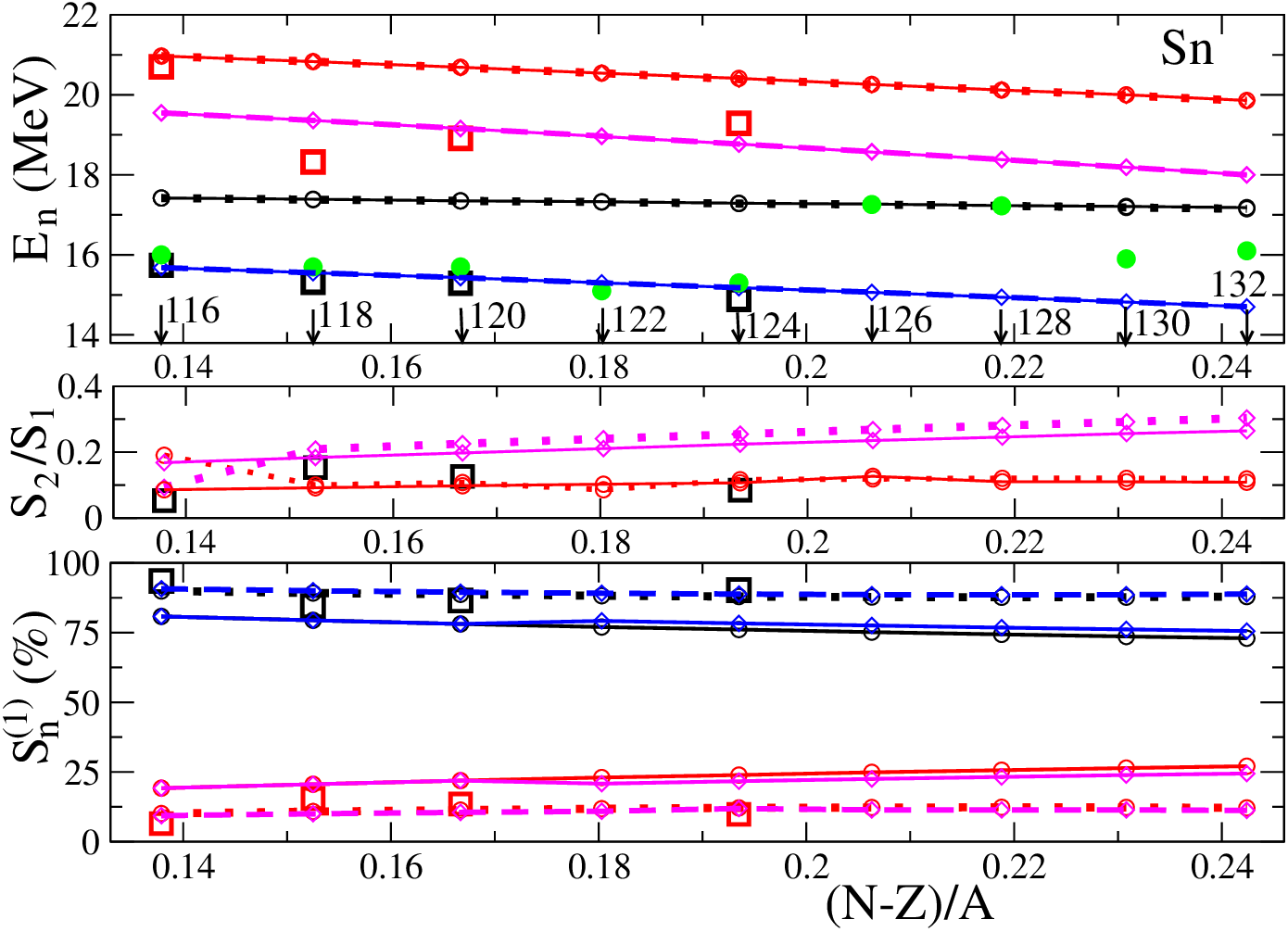}
\end{center}

\vspace{0.8cm}
\caption{(Color online) The IVDR splitting vs the asymmetry parameter 
$I=(N-Z)/A$.
{\it Top:} the energies $E_n$ of the main peak ($n=1$) and satellite ($n=2$);
black ($n=1$) and red ($n=2$) open squares are obtained from 
the experimental data for the integral cross sections of Refs.\ 
\cite{bermanSNexp,vanderwoudeSNexp,varlamovbook,varlamovprep,dietrichdata} 
for several Sn isotopes, as explained in the text \cite{kolmagsh}; 
the solid and dotted black lines with opened circles are the main L=50 MeV and 
0 peaks for SLy5* , respectively; the same red curves show the
satellites;
the solid and dashed blue lines with open diamonds denote the main L=60 MeV 
and 0
peaks for SVsym32* , respectively; the same margenta curves show the
satellites;
the green dots stands for  the averaged (IVGDR) experimental data  
(three last from
Adrich et al. \cite{adrich}) and arrows show the particle number of the Sn
isotopes. {\it Middle:} The ratio of the strengths at the satellite to 
those at the main peak, $S_n=S(\omega_n)$ 
(black solids and red dots for SLy5*; blue 
solids and margenta dashed curves
for SVsym32);
{\it Bottom:} The $S_n^{(1)}$ normalized to 100\% 
as explained in the text with the same notations as in the top plot.
}
\label{fig6}
\end{figure*}

\begin{figure*}
\begin{center}
\includegraphics[width=0.70\textwidth]{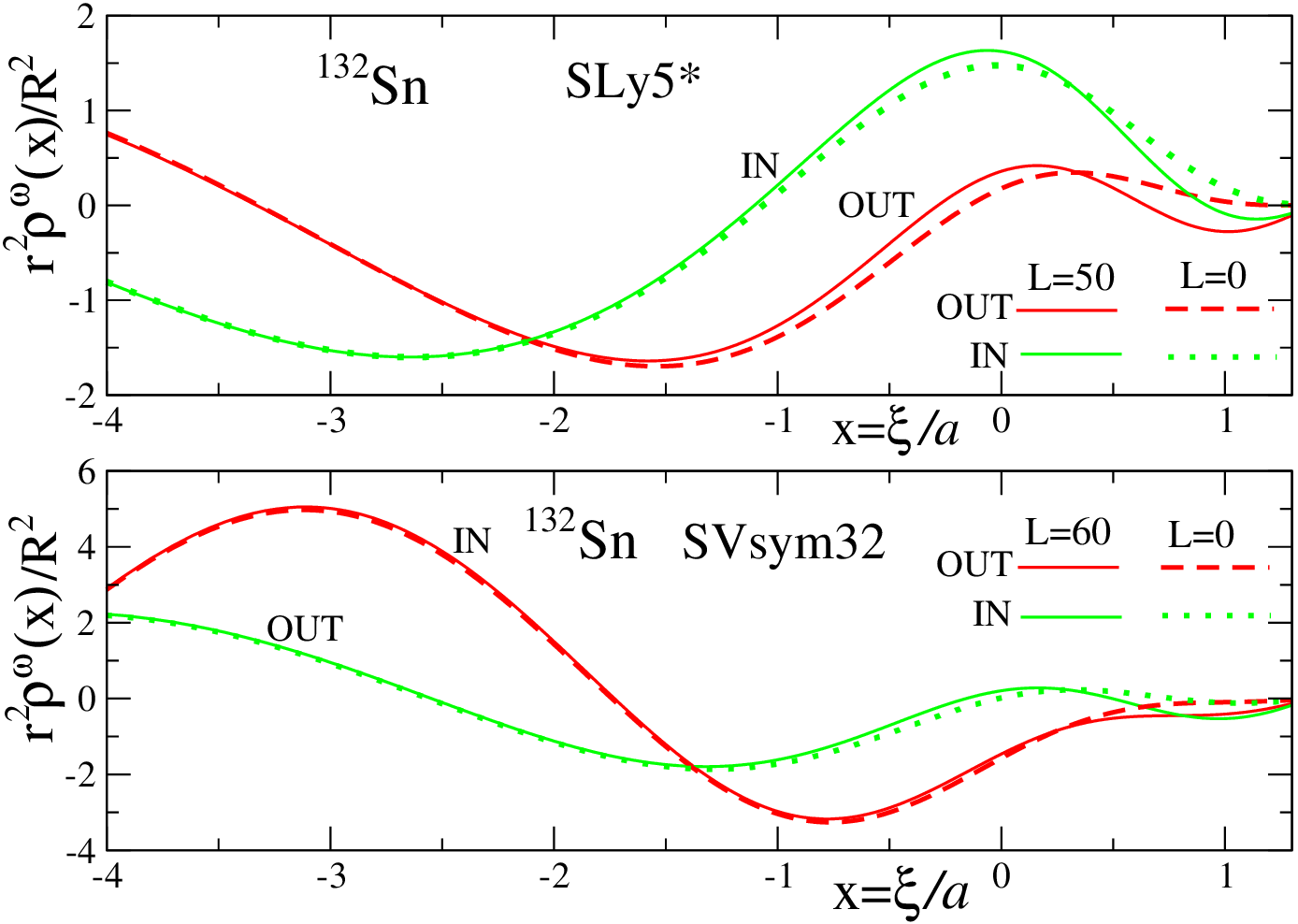}
\end{center}

\vspace{0.5cm}
\caption{(Color online)
The IVDR main out-of-phase ($\delta \rho_{-}$, ``out'') and
in-phase ($\delta \rho_{+}$,``in")  transition densities
$\rho_{-}^{\om}(x)$ [Eq.\ (\ref{drhom})]
 multiplied by $(r/R)^2$,
vs $x=\xi/a \approx (r-R)/a$ (spherical nuclei) for the satellite in $^{132}$Sn
with the Skyrme forces
 SLy5$^*$  \cite{pastore} (upper panel)
and SVsym32 \cite{reinhardSV} (lower panel);
the two characteristic values $L=0$ and $L=50$ (or $60$) MeV
are shown; the relaxation time $\mathcal{T}$
is the same as in Fig.\ \ref{fig3}.}
\label{fig7}
\end{figure*}

\begin{figure*}
\begin{center}
\includegraphics[width=0.70\textwidth]{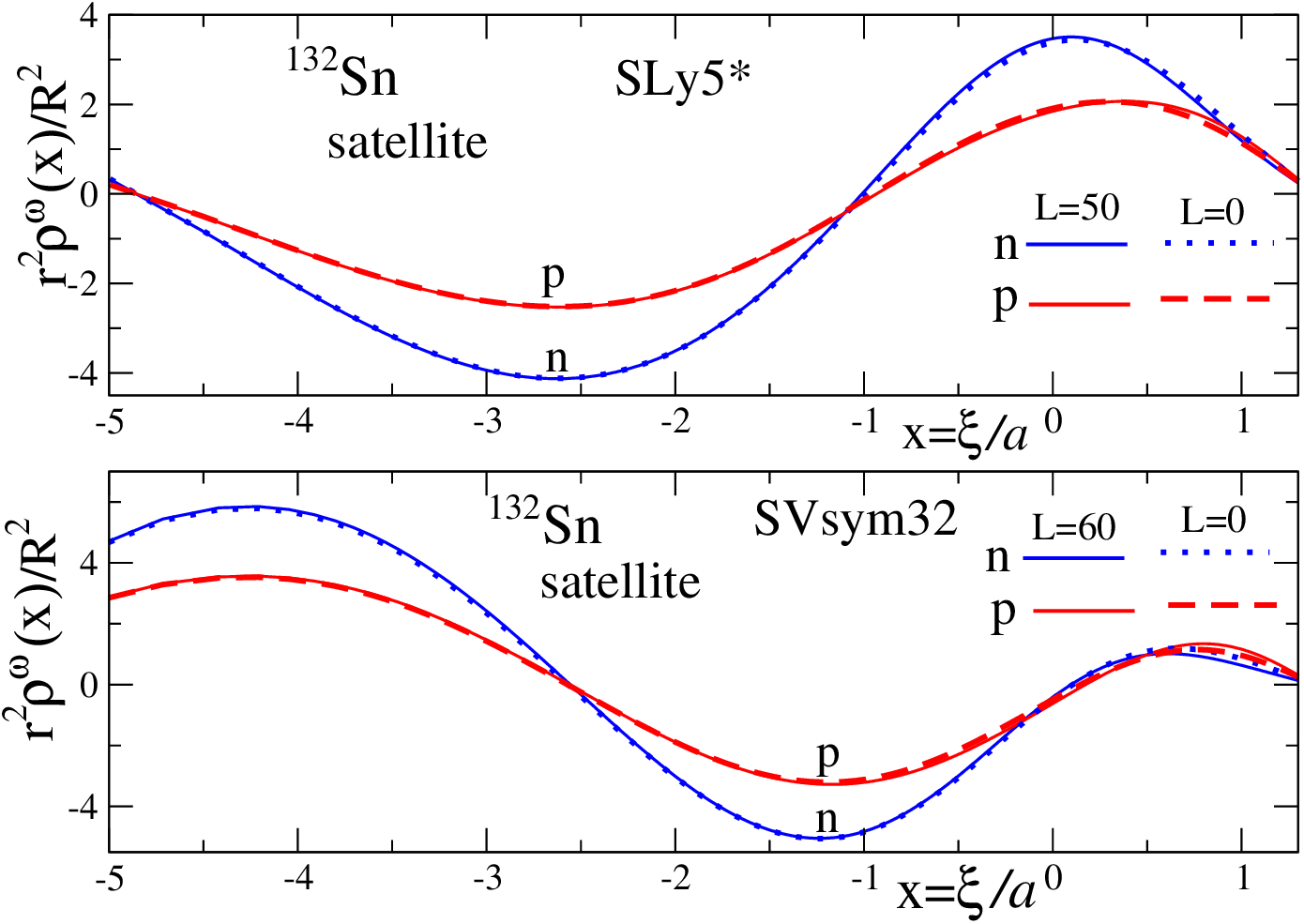}
\end{center}

\vspace{2.0cm}
\caption{(Color online) The same but for the IVDR neutron ($n$) and proton ($p$)
transition densities
$\rho^{\om}(x)$ [Eq.\ (\ref{drhom})]
multiplied by $(r/R)^2$,
vs $x=\xi/a \approx (r-R)/a$
for the satellite at the energy $E_2$ in $^{132}$Sn with the Skyrme forces
 SLy5$^*$  \cite{pastore,jmeyer} (upper panel)
and SVsym32 \cite{reinhardSV} (lower panel);
the two characteristic values $L=0$ and $L=50$ (or $60$) MeV
are shown too; the relaxation time $\mathcal{T}$
is the same as in Fig.\ \ref{fig3}.
}
\label{fig8}
\end{figure*}

\begin{figure*}
\begin{center}
\includegraphics[width=0.70\textwidth]{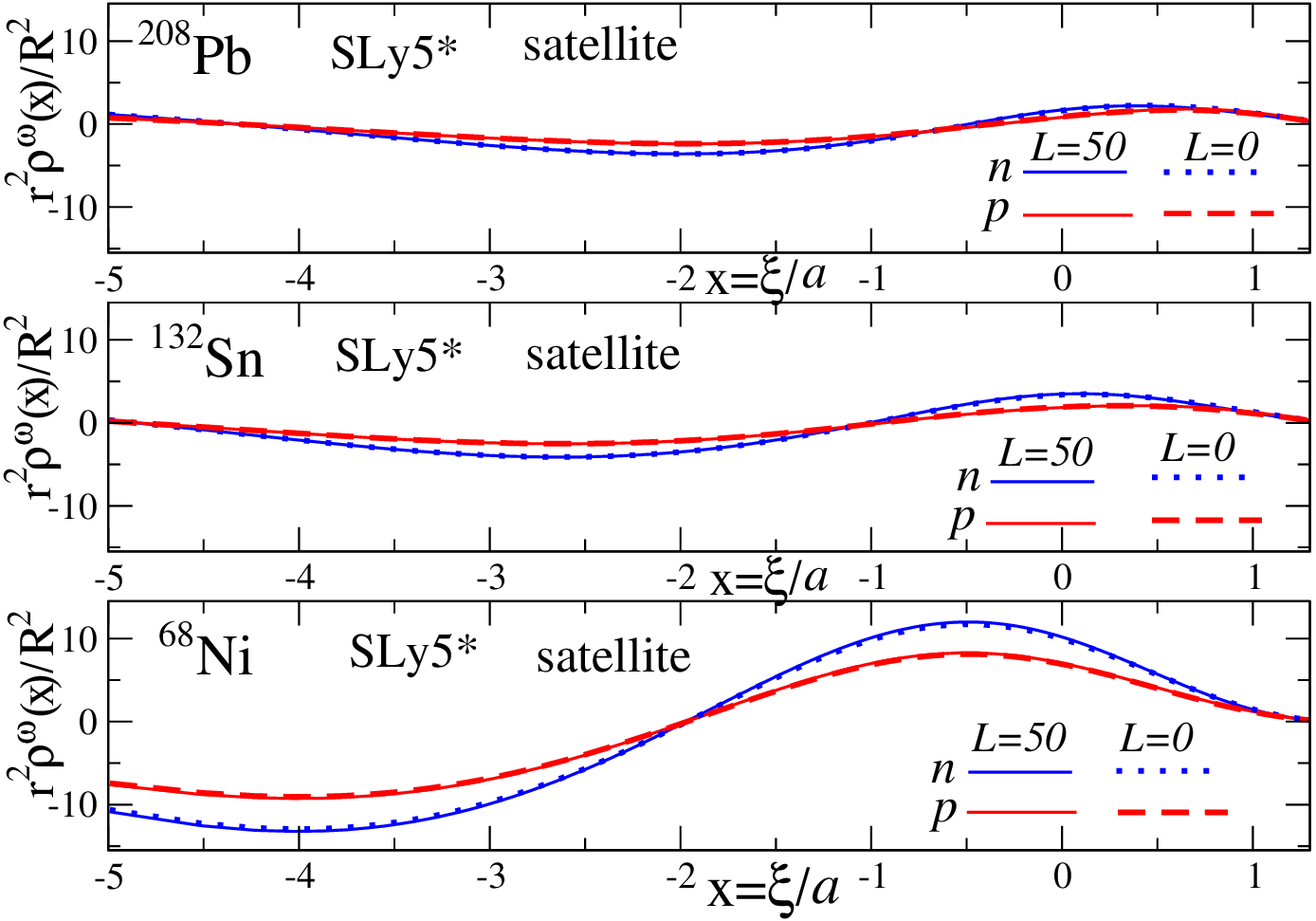}
\end{center}

\vspace{2.0cm}
\caption{(Color online)
The IVDR n-p transition densities $\rho^\omega(x)$ multiplied by
$(r/R)^2$ vs the dimensionless distance parameter $x=\xi/a \approx (r-R)/a$
for the same double magic nuclei slightly
depending on the slope parameter $L$ (in MeV) 
for a given example SLy5* of the Skyrme
forces as compared to the $L=0$ case near the ES edge
as in Fig.\ \ref{fig8}. }
\label{fig9}
\end{figure*}

\begin{figure*}
\begin{center}
\includegraphics[width=0.80\textwidth]{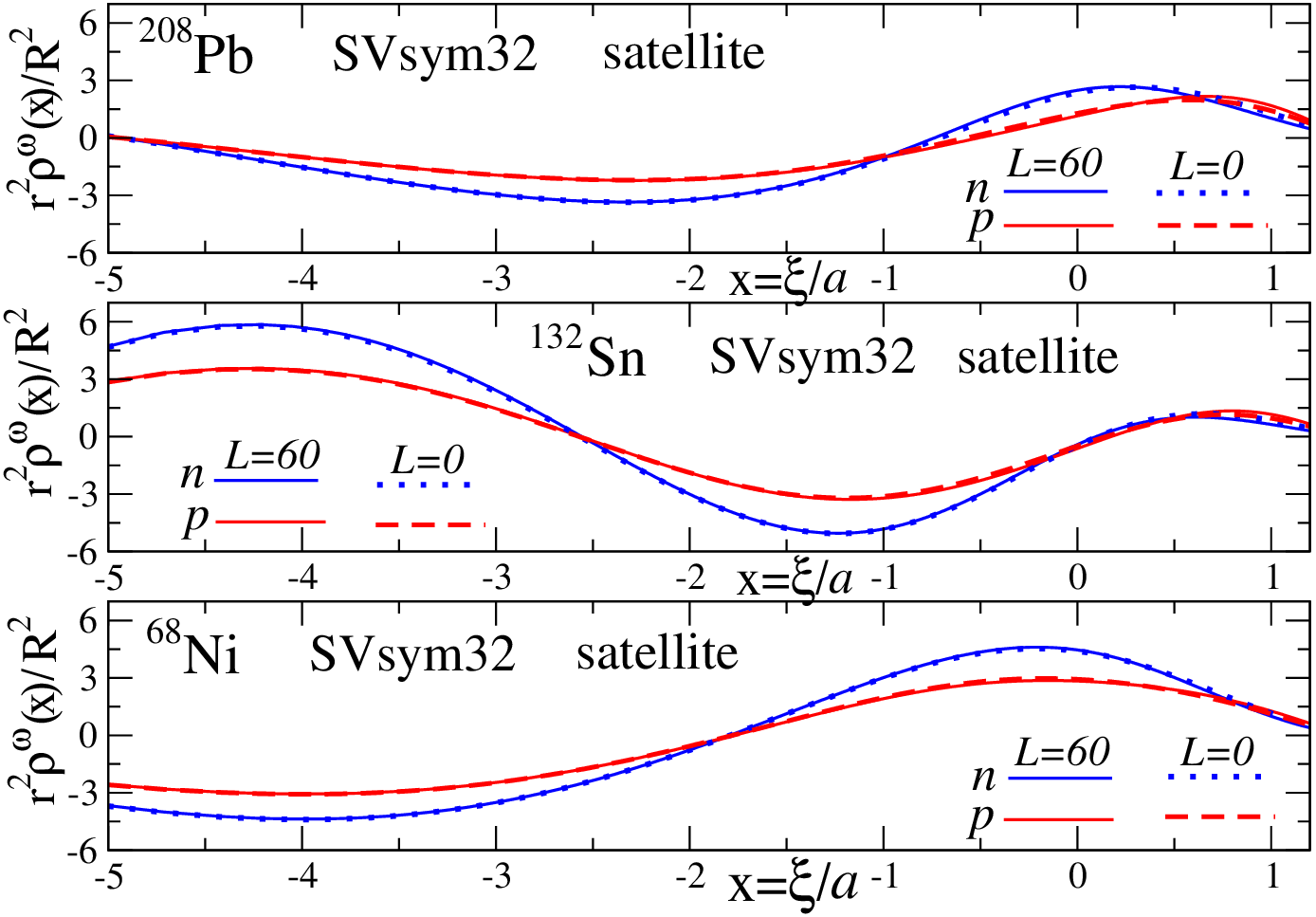}
\end{center}

\vspace{2.0cm}
\caption{(Color online) The same as in Fig.\ \ref{fig9} but for the Skyrme force SVsym32. }
\label{fig10}
\end{figure*}

\begin{figure*}
\begin{center}
\includegraphics[width=0.70\textwidth]{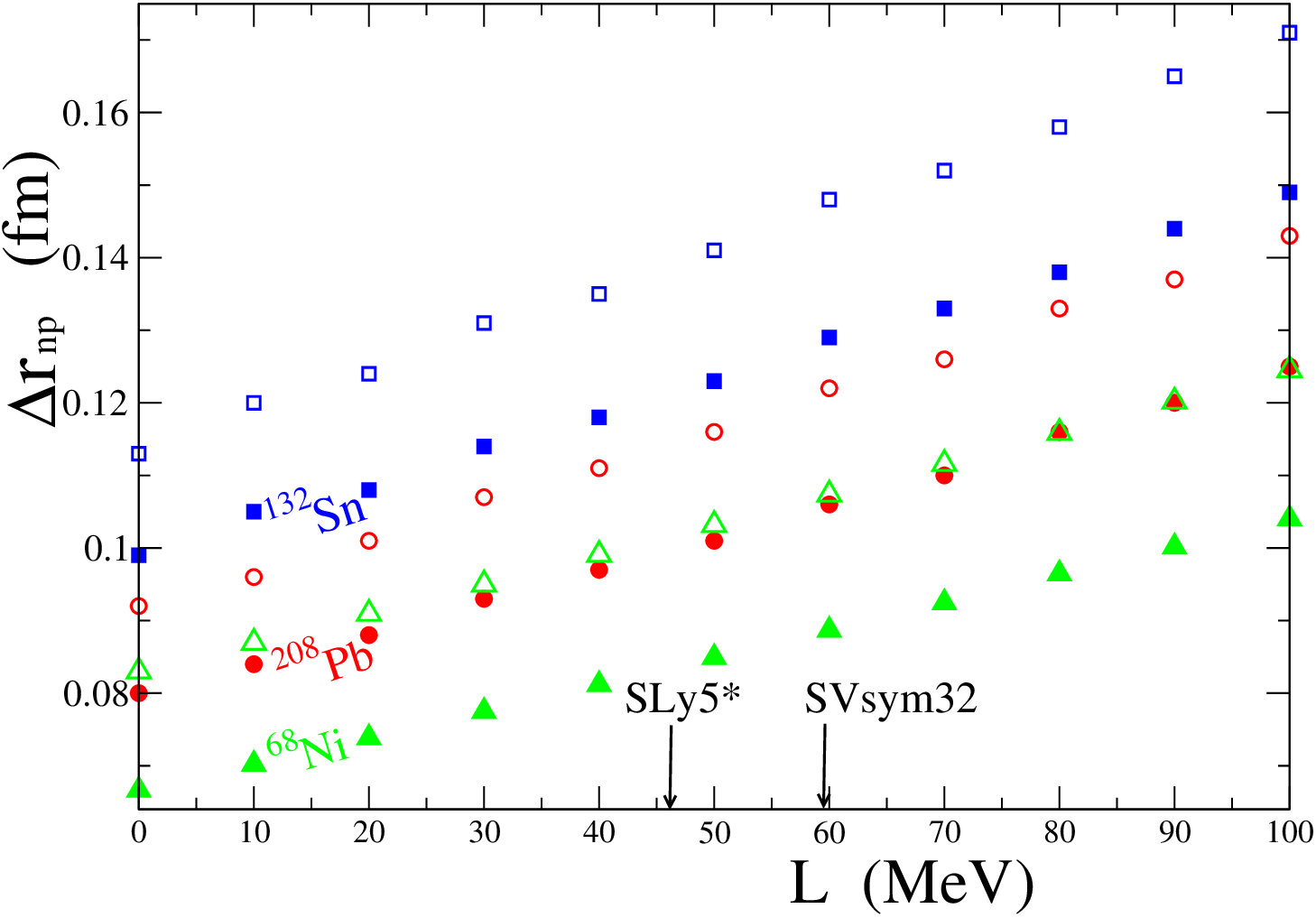}
\end{center}

\vspace{0.2cm}
\caption{(Color online) Neutron skin thickness
$\Delta r_{np}=\sqrt{3/5}\;(R_n-R_p)=\sqrt{3/5}\;r^{}_0 \tau$ ($~r_0=1.14$ fm)
as a function of the derivative constant $L$ for the same isotopes as in
Table II
and Figs.\ \ref{fig5}, \ref{fig9} and \ref{fig10} for
the SLy5* and SVsym32 forces; full symbols show SLy5* and open ones correspond
to SVsym32 calculations; arrows show approximately the values of $L$ for different Skyrme forces
taken from Refs.\ \cite{jmeyer,reinhardSV}.
}
\label{fig11}
\end{figure*}

The Lagrange equation for the variations of the isovector particle
density $\rho_{-}$ is given in the local coordinates $\xi,\eta$  by
\cite{BMRV,magsangzh}
\bea\l{lagrangeqminus}
&& 2 \mathcal{C}_{-}\frac{\partial^2 \rho_{-}}{\partial \xi^2 } + 2
\mathcal{C}_{-}\mathcal{H} \frac{\partial \rho_{-}}{\partial \xi}\nonumber\\
&-& \frac{\d}{\d \rho_{-}}\left[\rho_{+}
\varepsilon_{-}\left(\rho_{+},\rho_{-}\right)\right]
 + \lambda_{-}=0,
\eea
where $\mathcal{H}$ is the mean curvature of the ES, $\lambda_{-}$ is the
ES correction to the isovector chemical potential.
Up to the leading terms in a small parameter $a/R$ one gets from
Eq.\ (\ref{lagrangeqminus})
\be\l{lagrangeqminus0}
2 \mathcal{C}_{-} \; \frac{\partial^2 \rho_{-}}{\partial \xi^2} -
\frac{\d}{\d \rho_{-}} \left[\rho_{+} \epsd_{-}(\rho_{+},\rho_{-})\right]= 0\;.
\ee
We neglected here the higher order terms proportional to the first derivatives of the
particle density $\rho_{-}$ with respect to $\xi$ and
the surface correction to the isovector chemical potential in
Eq.\ (\ref{lagrangeqminus}) (Refs.\ \cite{strmagbr,strmagden} for
the isoscalar case).
For the dimensionless isovector density
$w_{-}=\rho_{-}/(\overline{\rho} I)$ one finds after simple transformations the
following equation and the boundary condition in the form
\bea\l{yeq0minus}
&& \frac{\d w_{-}}{\d w} =c_{\rm sym}
\sqrt{\frac{\overline{\mathcal{S}}_{\rm sym}(\eps)(1+ \beta w)}{e[\eps(w)]}}
\sqrt{\Big|1 -\frac{w_{-}^2}{w^2}\Big|},\\
&&
\quad w_{-}(w=1) =1\;,\nonumber
\eea
where $\beta$ is the SO parameter defined below Eq.\ (\ref{ysolplus}),
$\overline{\mathcal{S}}_{\rm sym}=\mathcal{S}_{\rm sym}/J$, $c_{\rm sym}$
is defined in Eq.\ (\ref{csymwt}) and
$\mathcal{S}_{\rm sym}(\eps)$ in Eq.\ (\ref{symen}).
The above equation determines the isovector density $w_{-}$ as a function of
the isoscalar one $w(x)$ [Eq.\ (\ref{ysolplus})]. In the
quadratic approximation for $e[\eps(w)]$ [up to a small asymmetry correction
proportional to $I^2$
in Eq.\ (\ref{epsilonplus})], one finds an explicit analytical expression
in terms of elementary functions \cite{BMRV}.
Substituting
$w_{-}=w\;\cos \psi$ into Eq.\ (\ref{yeq0minus}),
and taking the approximation $e=(1-w)^2$,
one has the following
first order differential equation for a new function $\psi(w)$:
\bea\l{ueqminus}
&-& \frac{w(1-w)}{c_{\rm sym}}\;\sin \psi\;\frac{\d \psi}{\d w} 
=\sqrt{\overline{\mathcal{S}}_{\rm sym}(\eps)(1+\beta w)}\; \sin\psi 
\nonumber\\
&-&
\frac{1-w}{c_{\rm sym}}\;\cos\psi,\qquad
\psi(w=1)=0.
\eea
The boundary condition for this equation is related to that
of Eq.\ (\ref{yeq0minus}) for $w_{-}(w)$.
This equation looks more complicated because of the
trigonometric nonlinear terms.
However, it allows one to obtain simple approximate
analytical solutions within standard perturbation theory.
Indeed, according to Eqs.\ (\ref{yeq0minus}) and (\ref{ysolplus}),
where we do not have an explicit $x$-dependence,
we note that $w_{-}$ is mainly a sharply decreasing function of
$x$ through $w(x)$ within a small diffuseness region of the
order of one in dimensionless units (Figs.\ \ref{fig1} and \ref{fig2}). Thus,
 we may find  approximate solutions to Eq.\ (\ref{ueqminus})
with its boundary condition in terms
of a power expansion of a new function
$\widetilde{\psi}(\gamma)$ in terms of a  new small
argument $\gamma$ [Eq.\ (\ref{csymwt})]
\be\l{powser}
\widetilde{\psi}(\gamma)\equiv
\psi(w)=\sum_{n=0}^{\infty} c_n\;\gamma^n(w)\;,
\ee
where the coefficients $c_n$ and $\gamma$ are defined in
Eq.\ (\ref{csymwt}). Substituting the power series
(\ref{powser}) into Eq.\ (\ref{ueqminus}), one expands first
the trigonometric functions into a power series of $\gamma$
according to the boundary condition in Eq.\ (\ref{ueqminus}).
As usual, using standard perturbation theory,
we obtain a system of algebraic equations
for the coefficients $c_n$ [Eq.\ (\ref{powser})] by
equating coefficients from both sides of
Eq.\ (\ref{ueqminus}) with the same powers of $\gamma$. This
simple procedure leads to a system of algebraic recurrence relations
which determine the coefficients $c_n$ as functions of the parameters
$\beta$ and $c_{\rm sym}$ of Eq.\ (\ref{ueqminus}),
\bea\l{cn}
c_0&=&0,\qquad c_1=\frac{1}{\sqrt{1+\beta}},\\
c_2&=&\frac{c_1}{2 c_{\rm sym}(1+\beta)}\left(
\beta c_{\rm sym}^2 + 2 + \frac{L}{3J} c_{\rm sym}^2(1+\beta)\right),
\nonumber\\
c_3&=&-c_1\left\{
\frac{4}{3} c_1^2 -3 \frac{c_1 c_2}{c_{\rm sym}} -
\frac{c_2 c_{\rm sym}}{2 c_1}\left(\beta c_1^2 + \frac{L}{3 J}\right) 
\right. \nonumber\\
&-& \left.\frac18 \beta^2 c_{\rm sym}^2 c_1^4
+ \frac{K_{-}c_{\rm sym}^2}{36 J}
+\frac{c_{\rm sym}^2L}{12 J} \left(\beta c_1^2  -
\frac{L}{6 J} \right)\right\},\nonumber
\eea
and so on. In particular, up to second order in $\gamma$,
we derive analytical solutions as functions of $\beta$, $c_{\rm sym}$, $J$
and $L$ in an explicitly closed form:
\bea\l{wsol3}
\widetilde{\psi}(\gamma) &=& \gamma\left(c_1 +
c_2 \gamma\right),\qquad
c_1=\frac{1}{\sqrt{1+\beta}},\\
c_2&=& \frac{\beta c_{\rm sym}^2 +2
+  L c_{\rm sym}^2 (1+\beta)/(3J)}{2 (1+\beta)^{3/2} c_{\rm sym}}.
\eea
Thus,  using the standard
perturbation expansion method of solving
$\widetilde{\psi}(\gamma)$ in terms of the power series
of the $\gamma$ up to $\gamma^2$, one obtains
the quadratic expansion of $\psi(w)$ [Eq.\ (\ref{ysolminus})]
with $\widetilde{c}=c_2/c_1$.
Notice that one finds a good convergence of the power expansion
of $\widetilde{\psi}(\gamma(w))$ (\ref{wsol3}) in $\gamma(w)$ for
$w_{-}(x)$ at the second order in $\gamma(w)$ because of values of
$c_{\rm sym}$ larger one for all Skyrme forces presented in Table I
[Eq.\ (\ref{csymwt}) for $c_{\rm sym}$].

\appendix \setcounter{equation}{0}
\renewcommand{\theequation}{B\arabic{equation}}

\begin{center}
\textbf{Appendix B: Derivations of the surface energy and
its coefficients}
\label{appC}
\end{center}

For the calculation of the surface energy components
$E_S^{(\pm)}$ of the energy $E$ in  Eq.\ (\ref{energy})
within the same improved ES approximation as described above
in Appendix A we first
separate the volume terms related
to  the first two terms of Eq.\ (\ref{enerdenin}) for the energy density
$\epsi$ per particle. Other terms of the energy density
 $\rho \epsi(\rho_{+},\rho_{-})$ in
Eq.\ (\ref{enerdenin}) lead to the surface components $E_S^{\pm}$
[Eq.\ (\ref{Espm})], as they
are concentrated near the ES.
Integrating the energy density $\rho \epsi$
per unit of the volume [see Eq.\ (\ref{enerdenin})]
over the spatial coordinates
$\r$ in the local coordinate system $\xi,\eta$ (see Fig.\ \ref{fig1}) in the ES
approximation, one finds
\bea\l{Espm1}
E_S^{\pm} &=&
\oint  \d S \int\limits_{\xi_{in}}^{\infty} \d \xi
\left[\mathcal{C}_{\pm}\left(\nabla \rho_{\pm}\right)^2
+ \rho_{+}\epsd_{\pm}\left(\rho_{+},\rho_{-}\right)\right]\nonumber\\
&\approx& \sigma_{\pm}\;S,
\eea
where $\xi_{in} \siml -a$
(Refs.\ \cite{strmagbr,strmagden,magsangzh}).
The local coordinates $\xi,\eta$
were used because the integral
over $\xi$ converges rapidly within the ES layer which is effectively taken
for $|\xi|\siml a$. Therefore again, we may extend
formally $\xi_{in}$ to $-\infty$ in the first (internal) integral
taken over the ES in the normal direction $\xi$ in
Eq.\ (\ref{Espm1}).
Then, the second integration is performed over the closed surface
of the ES. The integrand over $\xi$ contains terms of the order of
$(\overline{\rho}/a)^2 \propto (R/a)^2$ like the ones of the leading order
in the first equation of Ref.\ \cite{BMRV}.
However, the integration is effectively performed over the edge region
of the order of $a$ that leads to
the additional smallness proportional to $a/R$ like in Appendix A.
At this leading order the $\eta$ dependence of the internal
integrand can be neglected.
Moreover, from the Lagrange equations [see Eq.\ (\ref{lagrangeqminus0})
for the isovector case]
at this order one can realize that
terms without the
particle density gradients in Eq.\ (\ref{Espm1}) are equivalent to the gradient
terms. Therefore, for the calculation
of the internal integral we may approximately reduce the integrand
over $\xi$ to
derivatives of the universal particle densities
of the leading order $\rho_{\pm}(\xi)$ in $\xi$
using
$\mathcal{C}_{\pm} \left(\nabla \rho_{\pm}\right)^2 +
\rho_{+}\epsd_{\pm}\left(\rho_{+},\rho_{-}\right) \approx
2\mathcal{C}_{\pm}(\partial \rho_{\pm}/\partial \xi)^2$
[see Eqs.\ (\ref{ysolplus}) and (\ref{ysolminus}) for $w_{\pm}(x)$] .
We emphasize that the isovector gradient terms are obviously
 important for these calculations.
Taking
the integral over $\xi$ within the infinite
integration region ($-\infty <\xi <\infty$) out of the integral over
the ES ($\d S$) we are left with the
integral over the ES itself that is the
surface area $\mathcal{S}$. Thus,
we arrive finally at the right-hand side of Eq.\ (\ref{Espm1})
with the surface tension coefficient $\sigma_{\pm}=b^{(\pm)}_S/(4\pi r_0^2)$
[ see Eq.\ (\ref{sigma}) for $b^{(\pm)}_S$].

Using now the quadratic approximation $e[\eps(w)]=(1-w)^2$
in Eq.\ (\ref{sigma}) for $b_{S}^{\pm}$
($\mathcal{D}_{-} = 0$)
one obtains (for $\beta<0$, see Table I)
\be\l{bsJpm}
b_{S}^{(\pm)}= 6 \overline{\rho}\; \mathcal{C}_{\pm}\;
\frac{\mathcal{J}_{\pm}}{r_0 a}\;,
\ee
where
\bea\l{Jp}
\mathcal{J}_{+}&=&\int\limits_0^1 \d w\; \sqrt{w(1+\beta w)}\;(1-w)\\
&=&\frac{1}{24(-\beta)^{5/2}}\times\;\nonumber\\
&&\times\left[\mathcal{J}_{+}^{(1)}\; \sqrt{-\beta(1+\beta)}
+ \mathcal{J}_{+}^{(2)}\; \arcsin\sqrt{-\beta}\right],\nonumber
\eea
with
\be
\mathcal{J}_{+}^{(1)}=3 + 4 \beta(1+\beta),\quad\quad\quad
\mathcal{J}_{+}^{(2)}=-3-6\beta\;.
\ee
For the isovector energy constant $\mathcal{J}_{-}$ one finds
\bea\l{Jm}
\hspace{-2.0cm}\mathcal{J}_{-}&=&\frac{-1}{1+\beta}\;
\int\limits_0^1 \d w\; \sqrt{w(1+\beta w)}
(1-w)(1+\widetilde{c} \gamma(w))^2\nonumber\\
&=&\frac{\widetilde{c}^2}{1920 (1+\beta) (-\beta)^{9/2}}\;
\left[\mathcal{J}_{-}^{(1)}\left(\frac{c_{\rm sym}}{\widetilde{c}}\right)\sqrt{-\beta(1+\beta)}\right.\nonumber\\
&&\quad\quad\quad\quad\quad\quad\quad+\left.{\cal J}_{-}^{(2)}\left(\frac{c_{\rm sym}}{\widetilde{c}}\right)\arcsin\sqrt{-\beta}\right],
\eea
with
\bea
\mathcal{J}_{-}^{(1)}(\zeta)&=& 105- 4 \beta\left\{95 +75 \zeta +
\beta \left[119+10\zeta (19+6\zeta)\right.\right.\nonumber\\
 &+& \left.\left. 8 \beta^2
\left(1+ 10\zeta(1+\zeta)\right) + 8 \zeta \left(5 \zeta (3 +2 \zeta) -6\right)\right]\right\},
\nonumber\\
\mathcal{J}_{-}^{(2)}(\zeta)&=&15 \left\{7+2\beta \left[
5 (3 + 2 \zeta) + 8 \beta (1+\zeta)\right.\right.\nonumber\\
&\times&\left.\left.\left(3 +\zeta +2 \beta (1+\zeta)\right)\right]\right\}.
\eea
These equations determine explicitly the analytical expressions for the
isoscalar ($b_S^{(+)}$) and isovector ($b_S^{(-)}$) energy constants in terms of
the Skyrme force parameters; see
Eq.\ (\ref{ctilde}) for $\widetilde{c}$ and Eq.\ (\ref{csymwt})
for $c_{\rm sym}$
and $\gamma(w)$.
For the limit $\beta \rightarrow 0$ one has
from Eqs.\ (\ref{Jp}) and (\ref{Jm})
$\mathcal{J}_{\pm} \rightarrow 4/15$. With Eqs.\
(\ref{skin}) and (\ref{fw}) one arrives also
at the explicit analytical expression  for the isovector stiffness $Q$
as a function of $\mathcal{C}_{-}$ and $\beta$. In the limit
$\mathcal{C}_{-} \rightarrow 0$  one obtains $k^{}_S \rightarrow 0$ and
$Q \rightarrow \infty$ because of the finite limit of
the argument 
$c_{\rm sym}/\widetilde{c}\rightarrow 2(1+\beta)/[\beta +(1+\beta) L/(3J)]$
of the function $\mathcal{J}_{-}$ in Eq.\ (\ref{Jm})
[see also Eqs.\ (\ref{ysolminus}) for
$\widetilde{c}$ and Eq.\ (\ref{csymwt})
for $c_{\rm sym}$].

%%%%%%%%%%%%%%%%%%%%%%%%%%%%%%%%%%%

%\vspace{-10.0cm}
%
%TABLE I.
\noindent
\hspace{1cm}
\begin{table*}[pt]
\begin{tabular}{|c|c|c|c|c|c|c|c|c|c|c|c|}
\hline
 & SkM$^*$ & SGII & SLy5 &SLy5$^*$
& SLy6 & SLy7 & SVsym28 & SVsym32 &
SVmas08 & SVK226 & SVkap02\\
\hline
$\overline{\rho} $ (fm$^{-3}$)  & 0.16 &  0.16 & 0.16 & 0.16
  & 0.17 & 0.16 & 0.16 & 0.16 &
 0.16 & 0.16 & 0.16  \\
$b^{}_{V}$ (MeV) & 15.8&   15.6 & 16.0 & 16.0
   & 17.0 & 15.9 & 15.9 & 15.9 &
 15.9 & 15.9 & 15.9  \\
$K$ (MeV) & 217&   215& 230 & 230
   & 245 & 230 & 234 & 234 &
 234 & 226 & 234   \\
$J$ (MeV)  & 30.0 & 26.8 & 32.0& 32.0
& 32.0 & 32.0 & 28.0 & 32.0 &
 30.0& 30.0 & 30.0  \\
$L$ (MeV)  & 47.5 & 37.7 & 48.3& 45.9
& 47.4 & 47.2 & 7.5 & 59.5 &
 42.0& 35.5 & 37.0  \\
$\mathcal{C}_{+}$ (MeV$\cdot$fm$^{5}$) & 57.6 &   43.9 & 59.3 &60.1
 & 54.1 & 52.7& 49.6 & 51.8 &
 50.9& 51.4 & 50.7  \\
$\mathcal{C}_{-}$ (MeV$\cdot$fm$^{5}$) & -4.79 & -0.94 & -22.8  &-24.2
 & -15.6 & -13.4 & 19.6 & 26.0 &
 36.9 & 30.6 & 21.9 \\
$c_{\rm sym}$ & 3.24 & 6.07& 1.58 & 1.54
   & 1.77& 1.95 & 1.48 & 1.40 &
 1.13 & 1.22 & 1.46 \\
$\beta$ & -0.64 & -0.54 & -0.58 & -0.52
   &-0.62 & -0.65 & -0.48 & -0.47 &
 -0.51 & -0.48 & -0.48 \\
\hline
\end{tabular}

\vspace*{0.2cm}
TABLE\ I. Basic parameters of some critical Skyrme forces from
Refs.\ \cite{chaban,reinhardSV}, including the $L$ derivatives
\cite{vinas2,jmeyer, reinhardSV}. In addition to these standard quantities,
are
the isoscalar and isovector constants $C_{\pm}$ of the energy density gradient
terms [Eqs.\ (\ref{enerdenin}) and (\ref{Cpm})];
$c_{\rm sym}$ is given by Eq.\ (\ref{csymwt}) and the spin-orbit constant $\beta$
is defined below Eq.\ (\ref{ysolplus}).
\end{table*}

\vspace*{0.2cm}
%TABLE II
\noindent
\hspace{1cm}
\begin{table*}[pt]
\begin{tabular}{|c|c|c|c|c|c|c|c|c|c|c|c|}
\hline
 & SkM$^*$ & SGII & SLy5 &SLy5$^*$ &
SLy6 & SLy7 & SVsym28 & SVsym32 &
 SVmas08 & SVK226 & SVkap02\\
\hline
$k_{S,0}$(MeV)& -2.47 & -0.53 & -12.6 & -13.1 &
  -9.03 & -7.09 & 11.4 & 15.6 &
 37.1  & 23.7 & 12.7    \\
$k^{}_S$(MeV) & -2.48 &-0.46 & -14.6 & -15.0 &
  -10.1 & -7.61 & 13.3  & 18.2 &
 46.7  & 29.5 & 14.8  \\
$\nu^{}_0$  & 163 & 21.9 & 0.59 & 0.92 &
  1.21 & 1.99 & 0.90  & 0.84 &
 0.89  & 0.79 & 0.89   \\
$\nu$  & 2.27 & 1.89 & 0.28 & 0.60 &
  0.62 & 0.73& 0.58  & 0.61 &
 0.86  & 0.70 & 0.59    \\
$Q_0$(MeV)  &  59642 & 29908 & 73 & 72 &
 137 & 287 & -62 & -55 & -62  & -30 &
 -63     \\
$Q$(MeV)  &   823 & 2570 & 42 & 41 &
  63 & 98 & -34 & -34 & -34  & -21
 & -36   \\
$\tau^{}_0/I$ & 0.006 & 0.004 & 0.41 & 0.43
& 0.26 & 0.16 & 0.43 & 0.53 &
 0.040 & 0.89 & 0.45   \\
$\tau/I$ & 0.055 & 0.014 & 0.59 & 0.60 &
  0.40 & 0.28 & 0.62 & 0.73 &
 1.68 & 1.18 & 0.64  \\
$D_0$(MeV) $^{132}$Sn & 89  & 91 &
  101 & 89  & 104 & 102
   & 78 & 79 &
 81 & 77 & 84 \\
$D$(MeV) $^{68}$Ni & 91   & 92  &
100  & 88 & 104  & 95
   & 79  & 80  &
 83  & 78   & 85  \\
~~~~~~~~~~~~$^{132}$Sn & 89  & 91 &
100  & 89  &  103  & 95
   & 77 &  78  &
  81  & 76   &  83   \\
~~~~~~~~~~~~$^{208}$Pb &  90  &  91 &
109 & 88  & 102 & 93
   & 77 & 78 &
 81 & 76  & 82  \\
\hline
\end{tabular}

\vspace{0.2cm}
TABLE\ II. The isovector energy  $k^{}_S$
and the stiffness $Q$ coefficients are shown for several Skyrme forces
\cite{chaban,reinhardSV,jmeyer}; $\nu $ is the constant of Eq.\ (\ref{stiffin});
$\tau/I$ is the neutron skin thickness
calculated by Eq.\ (\ref{skin}) with the corresponding $L$ ; the functions
 $D(A)$ for the FLD model
in the last three lines
are calculated
with the relaxation time $\mathcal{T}$ having
the constant of its frequency dependence
$\mathcal{T}_{0 \rm Pb}=300$MeV$^2$$\cdot$ s
as explained in the text
and in the Figures
\cite{belyaev};
the quantities $k^{}_{S,0}$, $\nu^{}_0$, $Q_0$, $\tau^{}_0$
and $D_0$ are calculated with $L=0$.
\end{table*}

\newpage
%%%%%%%%%%%%%%%%%%%%%%%%%%%%%%%%%%%
% FIG.1:
%

%
%FIG. 2:

\newpage
%
% FIG.3
%

${}$
                \\[10.0ex]
\vspace{2.0cm}
%
% FIG.4
%
%

%
%FIG.5
%

%
%FIG. 6:

\vspace*{0.2cm}

\vspace{-50.5cm}

\vspace*{0.1cm}
%
% FIG.7
%

\vspace{-1.0cm}
%
% FIG.8
%
${}$
            \\[-10.0ex]

\newpage

%
%Fig.9
${}$
             \\[4.0ex]

%
%Fig.10

\clearpage
%
%Fig.11
%


\begin{thebibliography}{99}
%
\bibitem{strtyap} V.M. Strutinsky and A.\ S.\ Tyapin, JETP (Soviet Phys.)
{\bf 18}, 664 (1964).

%
\bibitem{strmagbr} V.M. Strutinsky, A.G. Magner, and M. Brack,  Z. Phys.
A {\bf 319}, 205 (1984).

%
\bibitem{strmagden} V.M. Strutinsky, A.G. Magner,
and V. Yu. Denisov, Z. Phys.
A {\bf 322}, 149 (1985).

%
\bibitem{brguehak} M. Brack, C. Guet, and H.-B. Hakansson,
Phys. Rep. {\bf 123}, 275 (1985).

%
\bibitem{magsangzh} A.G. Magner, A.I. Sanzhur, and A.M. Gzhebinsky,
Int. J. Mod. Phys. E {\bf 18}, 885 (2009).

%
\bibitem{BMV} J.P. Blocki, A.G. Magner, and A.A. Vlasenko, Nucl. Phys.
and At. Energy, {\bf 13}, 333 (2012).

%
\bibitem{BMRV} J.P. Blocki, A.G. Magner, P.Ring, and A.A. Vlasenko,
Phys.Rev. C {\bf 87}, 044304 (2013).

%
\bibitem{myswann69} W.D. Myers and W.J. Swiatecki,
Ann. Phys. (NY) {\bf 55}, 395 (1969); W.D. Myers and W.J. Swiatecki,
Ann. Phys. (NY)  {\bf 84}, 186 (1974).

%
\bibitem{myswnp80pr96} W.D. Myers, W.J. Swiatecki,
Nucl.Phys. A {\bf 336}, 267 (1980); W.D. Myers, W.J. Swiatecki,
Nucl.Phys. A
{\bf 601}, 141 (1996).

%
\bibitem{myswprc77} W.D. Myers et al., Phys. Rev. C {\bf 15}, 2032 (1977).

%
\bibitem{myswiat85} W.D. Myers, W.J. Swiatecki, and C.S. Wang,
Nucl.Phys. A {\bf 436}, 185 (1985).

%
\bibitem{vretenar1} D. Vretenar, N. Paar, P. Ring and G. A. Lalazissis,
Phys. Rev. C {\bf 63}, 047301 (2001).

%
\bibitem{vretenar2}
 D. Vretenar, N. Paar, P. Ring and G.A Lalazissis,
Nucl. Phys. A {\bf 692}, 496 (2001).

%
\bibitem{ponomarev} N. Ryezayeva, T. Hartmann, Y. Kalmykov et al.,
Phys. Rev. Lett., {\bf 89}, 272502 (2002).

%
\bibitem{danielewicz1} P. Danielewicz, Nucl. Phys. A {\bf 727}, 233 (2003).

%
\bibitem{pearson} M. Samyn, S. Goriely, M. Bender, and J.M. Pearson,
Phys. Rev. C, {\bf 70}, 044309 (2004).

%
\bibitem{danielewicz2} P. Danielewicz and J. Lee, Int. J. Mod. Phys.
E {\bf 18}, 892 (2009).

\bibitem{danielewicz3} P. Danielewicz, Nucl.Phys. A {\bf 818}, 36 (2009).

%
\bibitem{vinas1} M. Centelles, X. Roca-Maza, X. Vinas, and M. Warda,
Phys. Rev. Lett., {\bf 102}, 122502 (2009).

%
\bibitem{vinas2} M. Warda, X. Vinas, X. Roca-Maza, and M. Centelles,
Phys. Rev. C {\bf 80}, 024316 (2009); M. Warda, X. Vinas, X. Roca-Maza,
 and M. Centelles,
Phys. Rev. C {\bf 81}, 054309 (2010); M. Warda, X. Vinas, X. Roca-Maza, 
and M. Centelles,
Phys. Rev. C 
{\bf 82}, 054314 (2010).

%
\bibitem{vinas4}  X. Roca-Maza,  M. Centelles, X. Vinas, and M. Warda,
Phys. Rev. Lett. {\bf 106}, 252501 (2011).

%
\bibitem{kievPygmy} A. Voinov et al., Phys. Rev. C {\bf 81} 024319 (2010).
(2010).

%
\bibitem{larsenprc2013} A.C. Larsen et al., 
 Phys. Rev. C {\bf 87}, 014319 (2013).

%
\bibitem{nester1} A. Repko, P.-G. Reinhard, V.O. Nesterenko, and J.
Kvasil, Phys. Rev. C {\bf 87}, 024305 (2013).

%
\bibitem{nester2} W. Kleinig, V. O. Nesterenko, J. Kvasil, 
P.-G. Reinhard, and P. Vesely,
Phys. Rev. C {\bf 78}, 044313 (2008).

%
\bibitem{nester3} J. Kvasil, A. Repko, V.O. Nesterenko, W. Kleining,
P.-G. Reinhard,
Phys. Scr. T {\bf 154}, 014019 (2013).

%
\bibitem{vinas5}   X. Vinas,  M. Centelles, X. Roca-Maza,  and M. Warda,
Eur. Phys. J. A  {\bf 50}, 27 (2014).

%
\bibitem{Endressplitting} J. Endres et al., Phys. Rev. Lett.
{\bf 105}, 212503 (2010).


%
\bibitem{kolmag} V.M. Kolomietz and A.G. Magner, Phys. Atom. Nucl. {\bf 63},
1732 (2000).

%
\bibitem{kolmagsh} V.M. Kolomietz, A.G. Magner, and S. Shlomo,
Phys. Rev. C {\bf 73}, 024312 (2006).

%
\bibitem{BMRPhysScr2014} J.P. Blocki, A.G. Magner, and P. Ring,
Phys. Scr. T{\bf 89}, 054019 (2014).

%
\bibitem{adrich} P. Adrich et al., Phys. Rev. Lett., {\bf 95}, 132501 (2005).

%
\bibitem{reviewpygmy1} N. Paar, D. Vretenar, E. Khan, and G. Colo, 
Rep. Prog. Phys. {\bf 70}, 691 (2007).

%
\bibitem{wieland} O. Wieland  et al.,  Phys. Rev. Lett.,
  {\bf 102}, 092502 (2009).
%

\bibitem{reviewpygmy2} D. Savran, T. Aumann, and A. Zilges, 
Prog. Part. Nucl. Phys. {\bf 70},
210 (2013).

\bibitem{BMRPhysScr2015} J.P.\ Blocki, A.G.\ Magner, and P.\ Ring, 
 Phys. Scr. T {\bf 90}, 114009 (2015).

%
\bibitem{chaban} E. Chabanat et al.,
Nucl. Phys. A {\bf 627}, 710 (1997);  E. Chabanat et al.,
Nucl. Phys. A {\bf 635}, 231 (1998).

%
\bibitem{reinhard} P.-G. Reinhard and H. Flocard, Nucl. Phys. A {\bf 585},
467 (1995).

\bibitem{bender} M. Bender, P.-H. Heenen, P.-G. Reinhard,
Rev. Mod. Phys. {\bf 75}, 125 (2003).

\bibitem{revstonerein} J.R. Stone and P.-G. Reinhard, 
Prog. Part. Nucl. Phys. {\bf 58}, 587 (2007).

%
\bibitem{reinhardSV} P. Kl\"upfel P, P.-G. Reinhard, T.J. B\"urvenich, and
J.A. Maruhn, Phys. Rev. C {\bf 79}, 034310 (2009).
%

\bibitem{ehnazarrrein} J. Erler, C.J. Horowitz, W. Nazarevich, M. Rafalski, and
P.-G. Reinhard, arXiv:1211.6292v1 [nucl-th] (2012).

%
\bibitem{pastore} A. Pastore et al.,
Phys. Scr. T {\bf 154}, 014014 (2013).

%
\bibitem{jmeyer} J. Meyer (private communications, 2014).

%
\bibitem{PR2008}B.-A. Li, L.-W. Chen, C.M. Ko,
Phys. Rep. {\bf 464}, 113 (2008).

%
\bibitem{GU-HQ2013_PRC87-041301}
H. Q. Gu, H. Z. Liang, W. H. Long, N. V. Giai, and J. Meng,
Phys. Rev. C {\bf 87}, 041301 (2013).

%
\bibitem{bormot} Aa. Bohr and B. Mottelson, \textit{Nuclear Structure},
Vol. II (W.A. Benjamin, New York, 1975).

\bibitem{magstr} A.G. Magner and V.M. Strutinsky, Z. Phys. A 
   {\bf 322}, 633 (1985).

\bibitem{magboundcond} A.G. Magner,  Sov. J. Nucl. Phys. {\bf 45}, 235 (1987)
[Yad. Fiz. {\bf 45}, 374 (1987)].

\bibitem{denisov} V. Yu. Denisov, Sov. J. Nucl. Phys. {\bf 43}, 28 (1986).

%
\bibitem{belyaev} A.G. Magner, D.V. Gorpinchenko, and J. Bartel,
Phys. Atom. Nucl., {\bf 77}, 1229 (2014).

%
\bibitem{BM} J.P. Blocki and A.G. Magner,
Phys. Scr. T {\bf 154}, 014006 (2013).

%
\bibitem{eisgrei} J.M. Eisenberg, W. Greiner, {\it Nuclear Theory, Vol. I,
Nuclear Models Collective and Single-Particle Phenomena} (North-Holland,
 Amsterdam/London, 1970).

\bibitem{bermanSNexp} 
B.L. Berman and S.C. Fulz, Rev. Mod. Phys. {\bf 47}, 713
(1975).

\bibitem{vanderwoudeSNexp} A. Van der Woude, Prog. Part. Nucl. 
Phys. {\bf 18}, 217 (1987).

\bibitem{varlamovbook}  V.V. Varlamov, V.V. Sapunenko, 
and M.E. Stepanov, in 
{\it Photonuclear Data 1976-1995} (Moscow State University, Moscow, 1996), pp.\
1--220 [http://cdfe.sinp.msu.ru/service/index.html].

\bibitem{varlamovprep}  V.V. Varlamov, B.S. Ishanov, 
and M.E. Stepanov, 
{\it Systematics of Main Parameters of Atomic Nuclei Giant Dipole Resonances 
and Photonuclear Reaction Threshold Values} (Moscow State University, 
Institute of Nuclear Physics, Moscow, 1996). 

\bibitem{dietrichdata}  S.S. Dietrich and B.L. Berman, At. Data Nucl. 
 Data Tables {\bf 38}, 199 (1988).

%
\bibitem{rcnp} A. Tamii et. al., Phys. Rev. Lett. {\bf 107}, 062502 (2011).

%
\bibitem{gzhmagfed} A.M. Gzhebinsky, A.G. Magner, and 
S.N. Fedotkin, Phys. Rev. C {\bf 76}, 064315 (2007).

%
\bibitem{blmagyas} J.P. Blocki, A.G. Magner, and I.S. Yatsyshyn, Int. J.
Mod. Phys. E {\bf 21}, 1250034 (2012).

%
\bibitem{BMstatlev2012} J.P. Blocki and A.G. Magner, Phys. Rev. C {\bf 85},
064311 (2012).


\end{thebibliography}
\end{document}